\documentclass[journal]{IEEEtran}

\usepackage{fancyhdr} 
\usepackage{lastpage} 
\usepackage{extramarks} 
\usepackage[usenames,dvipsnames]{color} 
\usepackage{graphicx} 
\usepackage{listings} 
\usepackage{courier} 
\usepackage{lipsum} 
\usepackage{amsmath}
\usepackage{amssymb}
\usepackage{amsfonts}
\usepackage{amsthm}
\usepackage{cleveref}

\usepackage{autonum}

\def\eqdef{\stackrel{\triangle}{=}}

\newcommand{\permcap}{C_{\textsf{perm}}}
\newcommand{\rank}{\textsf{rank}}
\newcommand{\ext}{\textsf{ext}}

\newcommand{\az}{\ensuremath{k}}

\newcommand{\eqlinebreak}{\ensuremath{\nonumber \\ & \quad \quad}}

\newif\ifshorteq

\shorteqtrue

\ifshorteq

\newcommand{\eqlinebreakshort}{\ensuremath{\nonumber \\ & \quad \quad}}
\newcommand{\eqstartshort}{\ensuremath{&}}
\newcommand{\eqstartnotshort}{\ensuremath{}}
\newcommand{\eqbreakshort}{\ensuremath{ \\}}

\else

\newcommand{\eqlinebreakshort}{\ensuremath{}}
\newcommand{\eqstartshort}{\ensuremath{}}
\newcommand{\eqstartnotshort}{\ensuremath{&}}
\newcommand{\eqbreakshort}{\ensuremath{}}

\fi

\usepackage{custom_commands}
\usepackage{coloredboxes}
\title{Capacity of Noisy Permutation Channels}

\date{\today}

\author{Jennifer Tang,~\IEEEmembership{Member,~IEEE,} and Yury Polyanskiy,~\IEEEmembership{Senior Member,~IEEE,}
\thanks{This work was supported in part by the NSF grant CCF-2131115 and sponsored by the United States Air Force Research Laboratory and the United States Air Force Artificial Intelligence Accelerator and was accomplished under Cooperative Agreement Number FA8750-19-2-1000. The views and conclusions contained in this document are those of the authors and should not be interpreted as representing the official policies, either expressed or implied, of the United States Air Force or the U.S. Government. The U.S. Government is authorized to reproduce and distribute reprints for Government purposes notwithstanding any copyright notation herein.}
\thanks{J. Tang and Y. Polyanskiy are with the Department
of Electrical Engineering and Computer Science, Massachusetts Institute of Technology, Cambridge,
MA, 02139 USA e-mail: \{jstang,yp\}@mit.edu.}
\thanks{This paper was presented in part at ISIT 2022.}
}

\begin{document}

\maketitle

\begin{abstract} 
We establish the capacity of a class of communication channels introduced in~\cite{makur_2020}. The
$n$-letter input from a finite alphabet is passed through a discrete memoryless channel $P_{Z|X}$
and then the output $n$-letter sequence is uniformly permuted. We show that the maximal
communication rate (normalized by $\log n$) equals ${1\over 2} (\rank(P_{Z|X})-1)$ whenever
$P_{Z|X}$ is strictly positive. This is done by establishing a converse bound matching the
achievability of~\cite{makur_2020}. The two main ingredients of our proof are (1) a sharp bound on
the Kullback-Leibler divergence of a uniformly sampled vector from a type class and observed through a DMC to an iid vector; and (2)
the covering $\varepsilon$-net of a probability simplex with Kullback-Leibler divergence as a metric.  
In addition to strictly positive DMC we also find the noisy permutation capacity 
for $q$-ary erasure channels, the Z-channel and others.
\end{abstract}
 
\begin{IEEEkeywords}
Permutation channel, channel capacity, $\epsilon$-net covering
\end{IEEEkeywords} 
 
\section{Problem Statement and Main Results}


The noisy permutation channel, as formally introduced in~\cite{makur_2020}, is a communication
model in which an $n$-letter input undergoes a concatenation of a discrete memoryless channel
(DMC) and a uniform permutation of the $n$ letters. Since the receiver observes a uniformly permuted output, the order of symbols conveys no information. See
Section~\ref{sec:motiv} for a motivation of this model. More 
formally, the channel $P_{Y^n|X^n}$ can be described by the following Markov chain:
\begin{align}\label{eq::channel_diagram}
X^n \to Z^n \to  Y^n\,.
\end{align}
Here the channel input $X^n$ is a length $n$ sequence where each position takes a value in $\cX = [q]$ (where $[q] = \{1,2,\dots, q\}$). The sequence $X^n$ goes through the DMC which operates independently and identically on each symbol. This results in a sequence $Z^n$ where each position takes a value in $\cY = [\az]$. The DMC transition probabilities can be represented as a $q \times \az$ matrix $P_{Z|X}$. After the DMC, the sequence $Z^n$ goes through the permutation part of the channel and results in sequence $Y^n$ which is a uniformly random permutation of symbols on $Z^n$.

\begin{figure*}
    \centering
    \includegraphics[scale = .4]{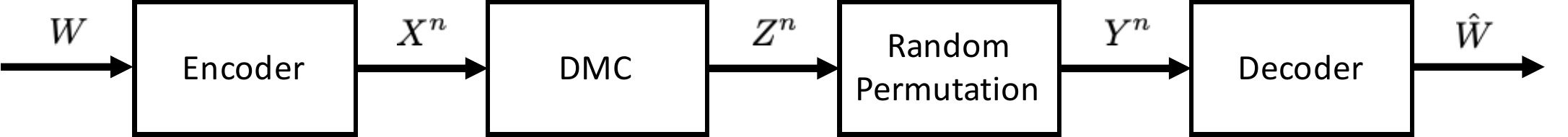}
    \caption{Diagram of the noisy permutation channel communication system. The key components are a DMC followed by a uniformly random permutation.}
    \label{fig::channel_diagram}
\end{figure*}

Let $f_n$ and $g_n$ be the channel encoder and decoder respectively. For each message $W \in [M]$, the input to the channel is $X^n = f_n(W)$. The output is $Y^n$, which the decoder decodes as $\hat W = g_n(Y^n)$. (See \Cref{fig::channel_diagram} for a diagram depicting the channel.) The probability of error is given by $P^{(n)}_{\textsf{error}} \eqdef \bbP[W \neq \hat W].$
The rate\footnote{Notice that rate $R$ for the noisy permutation channel is not the commonly used definition where $R = \frac{\log M}{n}$. The noisy permutation channel would have rate $0$ under this commonly used definition. Defining rate as in \eqref{eq::rate_def} is appropriate given that we intend to find capacity. Let $M^*(n, \epsilon) = \max\{M: \exists (n, M, \epsilon)-\text{code} \}$ where $n$ is the length (or channel uses), $M$ is the message size, and $\epsilon$ is the error probability for a given code (see \cite{polyanskiy2014lecture} for more details). Capacity is defined as the limit as $\epsilon \to 0^+$ of the coefficient of the leading term of $\log M^*(n, \epsilon)$. In the case of the noisy permutation channel, the leading term of $\log M^*(n, \epsilon)$ scales as $\log n$.
} for the encoder-decoder pair $(f_n, g_n)$ is defined as
\begin{align}\label{eq::rate_def}
R \eqdef \frac{\log M}{\log n}\,.
\end{align}
A rate $R$ is \emph{achievable} if there is a sequence of encoder-decoder pairs $(f_n, g_n)$ with rate $R$ such that $\lim_{n \to \infty} P^{(n)}_{\textsf{error}} = 0$.
The capacity for the noisy permutation channel with DMC $P_{Z|X}$ is
$\permcap(P_{Z|X}) \eqdef \sup \{R \geq 0: R \text{ is achievable} \}\,.$

In \cite{makur_2020}, the author determined that the noisy permutation channel capacity\footnote{While it might seem that the noisy permutation channel capacity should be a continuous function of the values in $P_{Z|X}$, note that this is not the case due to how capacity is defined. Changing values in $P_{Z|X}$ by a small $\delta$ could change the rank of $P_{Z|X}$ by $1$, but no matter how small $\delta$ is, there exists an $n$ large enough so the effects of $\delta$ can make a difference. } for DMC $P_{Z|X}$ is bounded by 
\begin{align}\label{eq::anuran_ach}
\permcap(P_{Z|X}) \geq \frac{\rank(P_{Z|X}) - 1}{2}\,.
\end{align}
 For strictly positive matrices $P_{Z|X}$ (meaning all the transition probabilities are greater than $0$), the author shows a converse bound 
\begin{align}
\permcap(P_{Z|X}) \leq \frac{|\cY| - 1}{2}\,.
\end{align}
 
 The author also gives a second converse bound: $\permcap(P_{Z|X}) \leq ({\ext(P_{Z|X}) - 1})/2$, where $\ext(P)$ is the number of extreme points of the convex hull of the rows of $P$. For the case of strictly positive DMC $P_{Z|X}$, these upper and lower bounds do not necessarily match if the rank of matrix $P_{Z|X}$ does not equal to $|\cY|$ or $\ext(P_{Z|X})$.

\subsection{Main Results}

Our main result is establishing tightness of the lower bound~\eqref{eq::anuran_ach}, resolving
Conjecture 1 of \cite{makur_2020}.
\begin{ftheorem}[Strictly Positive DMC]\label{thm::strict_positive}
For strictly positive $P_{Z|X}$,
\begin{align}
\permcap(P_{Z|X}) = \frac{\rank(P_{Z|X}) - 1}{2}\,.
\end{align}
\end{ftheorem}

Our proof uses the idea of covering the space of distributions via an $\varepsilon$-net under the
Kullback-Leibler (KL) divergence as a ``distance''\footnote{KL divergence is not technically a distance or metric (as it is not symmetric and does not follow triangle inequality), but we choose to use the term \emph{distance} since we are using KL divergence to measure how far two probability distributions are.}, following upon our investigations of a similar question
in~\cite{adler2021quantization}. In order to reduce to the covering question, we first need
another result that is, perhaps, of separate interest as well.

Let $\az$ be the alphabet size. 
We let $\cP_n$ be the set of $n$-types (probabilities which can be written as rational numbers with denominator $n$), i.e.
\begin{align}
    \cP_n = \bigg\{ P &\in \Delta_{\az - 1}: P = \left(\frac{a_1}{n}, \dots, \frac{a_\az}{n} \right) 
    \\&\text{ where } a_1,\dots, a_\az \in \bbZ_{\geq 0} \bigg\}\,.
\end{align}
(We use $\Delta_{\az - 1}$ for the $k-1$ dimensional probability simplex, see \Cref{sec::notation}.)
For $P \in \cP_n$, let $T_n(P)$ be the set of sequences of length $n$ in the type class\footnote{See Section 11.1 of \cite{coverthomas} for more background on types.} of $P$, i.e.
\begin{align}
    T_n&(P) = \Bigg\{s_1\dots s_n: s_t \in [\az]
    \\&
    \text{ and } \left(\frac{\sum_{t = 1}^n \bbI\{ s_t = 1\}}{n}, \dots , \frac{\sum_{t = 1}^n \bbI\{ s_t = \az\}}{n} \right) = P \Bigg\}
\end{align}
where $\bbI\{\cdot\}$ is the indicator function.
The notation $Q_Y$ means a distribution on random variable $Y$. For any distribution $Q_Y$, we use $Q_Y^n$ to mean the product distribution $Q_Y^n(y^n) = \prod_{t = 1}^n Q_Y(y_t)$. 
For any distribution $U$ on length $n$ sequences, the distribution $P_{Y|X}^n \circ U$ can be understood as the distribution on random sequences derived by first randomly selecting a sequence according to $U$, then passing each symbol in this sequence through the transition probabilities $P_{Y|X}$ independently. (See \Cref{sec::notation} for more discussion.)

\begin{figure}
\centering
\includegraphics[scale = .3]{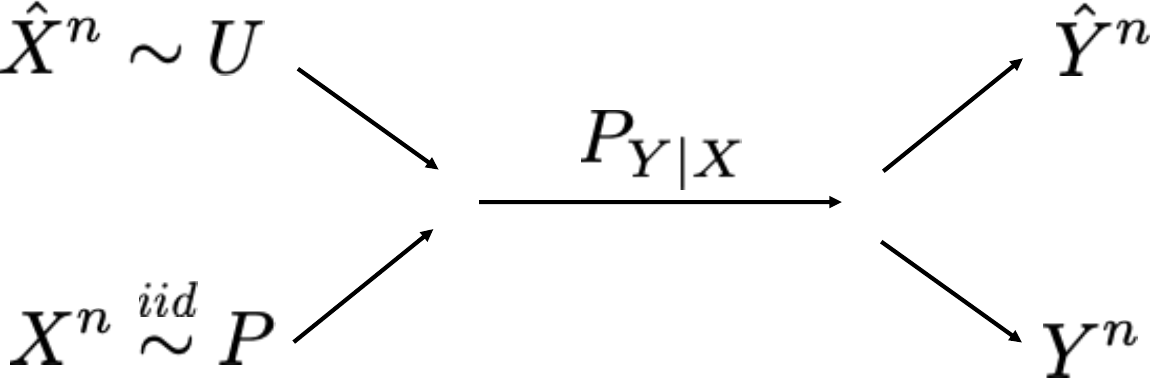}
\caption{ This diagram illustrates the  special case of \Cref{thm::bounds_uniform_type} where $Q_Y = P_Y$. Variable $X^n$ is distributed iid according to $P$ whereas $\hat X^n$ is a sequence uniformly drawn from type class $T_n(P)$ (a distribution represented by $U$). Variables $Y^n$ and $\hat Y^n$ are noisy versions of $X^n$ and $\hat X^n$ respectively.  \label{fig::compare_uniform_iid_Y}}
\end{figure}

Our next result deals with the following scenario: 
Select some $P \in \cP_n$ and
suppose we have two sequences, $X^n$ and $\hat X^n$. The sequence $X^n$ is generated iid using the
probability $P$. On the other hand, $\hat X^n$ has uniform probability over all sequences in the
type $T_n(P)$. Both sequences $X^n$ and $\hat X^n$ undergo the transition $P_{Y|X}$ applied
independently on each symbol and respectively results in $Y^n$ and $\hat Y^n$. How different are
the distributions of $Y^n$ and $\hat Y^n$ under KL divergence? (See
\Cref{fig::compare_uniform_iid_Y} for a diagram representing the relations of these variables.) Another interpretation of this
scenario is if there are $n$ balls of $q$ colors in an urn. The sequence $X^n$ are $n$ draws from the urn with
replacement and $\hat X^n$ are $n$ draws without replacement (in which case all the balls are drawn). These
observations then both go through the same noisy process to produce $Y^n$ and $\hat Y^n$.

It turns out that if $P_{Y|X}$ is strictly positive, then regardless of the sequence length $n$, 
\begin{align}
D(P_{\hat Y^n} \| P_{Y^n}) \leq c
\end{align}
where $c$ is a constant that only depends on $P_{Y|X}$. Our next result will actually
show something more general. The sequence $X^n$ can be generated iid with another distribution $Q$, and the KL divergence can still be bounded by constant $c$ plus another term which is the KL divergence of the marginals on $Y$ generated by $P$ and $Q$.

\begin{ftheorem}\label{thm::bounds_uniform_type}
Fix channel $P_{Y|X}$, where  $P_{Y|X}$ is strictly positive. Then there exists a constant $c = c(P_{Y|X})$ such that the following holds: For any $n$-type $P \in \cP_n$, let $U$ be uniform on $T_n(P)$. For all $Q_Y$ we have
\begin{align}\label{eq::bounds_uniform_type_eq}
n D(P_Y \| Q_Y) \leq D(P_{Y|X}^n \circ U \| Q_Y^n) \leq n D(P_Y\|Q_Y) + c
\end{align}
where $P_Y$ is the marginal distribution of $Y$ under $(P \times P_{Y|X})$.
\end{ftheorem}

\begin{remark}\label{rem::constant_c}
It can be shown that the constant $c$ in \Cref{thm::bounds_uniform_type} is
\begin{align}\label{eq::c_value}
c \leq \frac{q-1}{2} \log \frac{2 \pi \alpha^2}{c_{*}} + \frac{q}{12 n}\leq \frac{q-1}{2} \log \frac{2 \pi \alpha^2}{c_{*}} + \frac{q}{12}
\end{align}
where $\alpha$ is a universal constant defined in \Cref{thm::petrov_sum} (see \Cref{sec::petrov_concentration}) and if $p_{bj}$ denote the values in matrix $P_{Y|X}$,
\begin{align}
c_* = \min_b \frac{\min_j p_{bj}}{\max_j p_{bj}}\,.
\end{align}
\end{remark}

\begin{figure}
    \centering
    \includegraphics[scale = .55]{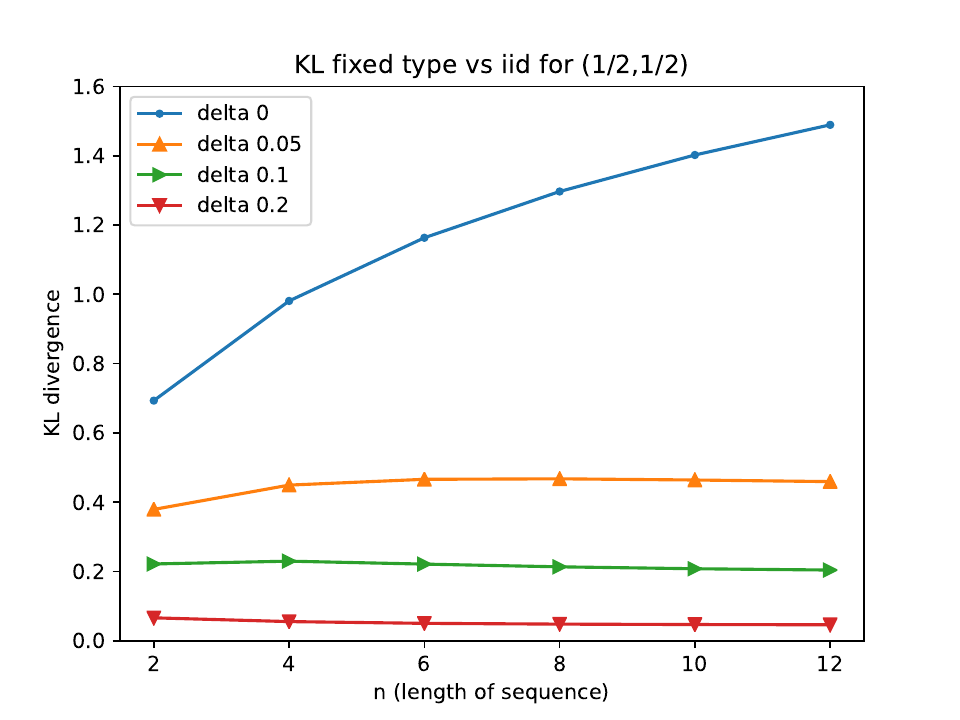}
    \caption{This plot demonstrates the consequences of \Cref{thm::bounds_uniform_type}. We numerically compute $D(P_{\hat Y^n} \| P_{Y^n})$ when $P = (1/2, 1/2)$ and  $P_{Y|X}$ is given as in \eqref{eq::bsc_example_transition} for different values of $\delta$. When $\delta = 0$, we see that the KL divergence is trending towards infinity (specifically it grows as $\frac{1}{2} \log n$). When any noise is added ($\delta$ is positive), this growth to infinity completely disappears. The quantity $D(P_{\hat Y^n} \| P_{Y^n})$ becomes constant in $n$, as \Cref{thm::bounds_uniform_type} states. }
    \label{fig::KL_noise_experiment}
\end{figure}

It is necessary in \Cref{thm::bounds_uniform_type} that $P_{Y|X}$ is strictly positive. In fact, it is surprising that \Cref{thm::bounds_uniform_type} can show that the KL divergence of $D(P_{Y|X}^n \circ U \| Q_Y^n)$ when $Q_Y = P_Y$ is constant, considering that this is not the behavior we would expect for transitions $P_{Y|X}$ which are not strictly positive. For example, using our simpler initial example with $X, \hat X, Y, \hat Y$ (which is depicted by \Cref{fig::compare_uniform_iid_Y}), consider when $X, Y \in [2] =  \{1,2\}$ and %
\begin{align}
P_{Y|X} = \begin{cases}
1-\delta & \text{ if } X = Y \\
\delta & \text{ if } X \neq Y 
\end{cases}\label{eq::bsc_example_transition}
\end{align}
This transition probability represents a binary symmetric channel (BSC) with crossover probability $\delta$. Suppose that $X \in T_n((1/2, 1/2))$. If $\delta = 0$ (and thus $P_{Y|X}$ is not strictly positive), then we can compute that 
\begin{align}
D(P_{\hat Y^n} \| P_{Y^n}) \approx \frac{1}{2}\log n\,.\label{eq::bsc_no_noise_KL}
\end{align}
However, if we increase $\delta$ slightly (adding any amount of crossover noise), this completely eliminates the growth of $D(P_{\hat Y^n} \| P_{Y^n})$ in $n$. For any positive $\delta$, $D(P_{\hat Y^n} \| P_{Y^n})$ is constant as $n$ increases, where the value of the constant depends on $\delta$. An illustration of this example is given in \Cref{fig::KL_noise_experiment}.

We note also that \Cref{thm::bounds_uniform_type} implies that the divergence of (a complicated distribution) $P_{\hat Y^n}$
to any iid distribution $Q^n_Y$ can be approximated with $nD(P_Y \| Q_Y)$ and this approximation will only be off by an additive
constant.

%

\begin{remark}\label{rem::stam_urn}

Note that  $D(P_{\hat
X^m}\| P_X^m)$ describes the difference between sampling $m$ balls from an $n$-urn with and
without replacement.  This is a classical question studied in \cite{stam1978}. Our setting studies this question for the particular case when $n = m$ and when the observations are noisy. Bounds for the noiseless case $D(P_{\hat
X^m}\| P_X^m)$ can still be an upper bound for the noisy case if we apply the data processing inequality. This shows that $D(P^n_{Y|X} \circ U \| P_Y^n) \leq D(P_{\hat
X^n} \| P_X^n) \leq
\frac{\az-1}{2}(\log n + c)$, where the second inequality is shown using Stirling's approximation. Our result removes the $\log n$ term in this bound, but only
under the assumption of a strictly positive $P_{Y|X}$. See \Cref{sec::comparing_stam} for more details on comparing our bound to that of \cite{stam1978} when $m < n$. We also note that results of
\cite{stam1978} as shown in
\cite{diaconis_freedman1980} imply the finitary case of de Finetti's theorem.
\end{remark}

Other contributions of this work use similar techniques to get converse results in other settings which do not have strictly positive DMC matrices. 

\begin{ftheorem}\label{thm::other_channels}
Other channel results:
\begin{enumerate}
\item \label{thm::block_diagonal} Suppose $P_{Z|X}$ can be written as a block diagonal matrix with $\beta$ blocks where each block is strictly positive. Then,
\begin{align}\label{eq::block_diagonal}
\permcap(P_{Z|X}) = \frac{\rank(P_{Z|X}) + \beta - 2}{2}\,.
\end{align}
\item \label{cor::q_erasure} For DMC $P_{Z|X}$ which is a q-ary erasure channel for $q \geq 2$ (assuming erasure probabilities are not $0$ or $1$), then 
\begin{align}
\permcap(P_{Z|X}) = \frac{q - 1}{2}\,.
\end{align} 
\item \label{cor::z_cap} For DMC $P_{Z|X}$ which is a Z-channel, then
\begin{align}
\permcap(P_{Z|X}) = \frac{1}{2}\,.
\end{align} 
\end{enumerate}
\end{ftheorem}

The first result in \Cref{thm::other_channels} applies to DMC $P_{Z|X}$ which are block diagonal matrices where each block is strictly positive. As \eqref{eq::block_diagonal} of \Cref{thm::other_channels} implies, we are able to show both the achievability and converse results for block diagonal DMC matrices. We prove both these results in \Cref{sec::block_diagonal}. This result also immediately illustrates that \Cref{thm::strict_positive} without the strictly positive condition cannot be true, since block diagonal matrices with $2$ or more strictly positive blocks violate the bound in \Cref{thm::strict_positive} entirely.

The second result is for binary erasure channels and $q$-ary erasure channels. The work in \cite{makur_2020} determines the capacity for when the DMC matrix is the binary symmetric channel (BSC), but leaves the binary and $q$-ary erasure channels as open problems. \Cref{cor::q_erasure} of \Cref{thm::other_channels} resolves Conjecture 2 presented in \cite{makur_2020} regarding the capacity of binary erasure channels and the conjecture regarding $q$-ary erasure channels\footnote{Note that while binary erasure channels and $q$-ary erasure channels usually have the same erasure probability for each symbol, \Cref{cor::q_erasure} of \Cref{thm::other_channels} is still true even if these erasure probabilities are different. The only requirement is that none of the erasure probabilities are $0$ or $1$. The capacity when the erasure probabilities are $0$ or $1$ is discussed in \cite{makur_2020}. }. 
This result uses \eqref{eq::anuran_ach} which is proved in \cite{makur_2020} as the achievability. Our contribution is the tight converse argument.

The third result in \Cref{thm::other_channels} deals with DMC which is the Z-channel. While this is tight, if we generalize to a $q$-ary Z-channel (or what we call in this work a ``zigzag'' channel), we are not always able to find tight results with our current covering technique. The erasure channels, Z-channels, and a brief analysis on the zigzag channel are discussed in \Cref{sec::other_channels}. 

All of these results also use the method of covering. A \emph{covering} is a set of points in a space (we call them centers) for which all other points in the space are within a certain distance $\varepsilon$ to (see \Cref{def::covering}). Using covering as a technique to determine the capacity for the noisy permutation channel is reasonable because the centers which are far apart can intuitively be equated with messages that are distinguishable. When the messages correspond to two distributions $Q_1$ and $Q_2$ which are far in KL divergence, it is unlikely that noisy versions of $Q_1$ will be close to noisy versions of $Q_2$. If two distributions are close in KL divergence, their noisy versions are likely to be confused. If the messages in our communication are centers of a covering, then we know that if we add another center (or message), it will be close to one of the existing covering centers and thus cause error in determining which of the centers (or messages) was sent. This gives us a limit on the total number of messages which can be sent, creating a converse bound\footnote{A similar notion to covering is packing, which is a set of centers in a space where all the centers in the set are at least distance $2\varepsilon$ from another. Intuitively, covering corresponds to a converse bound while packing corresponds to an achievabaility bound.}.

In order to use this intuition mathematically, we need to overcome the obstacle of computing the KL divergence over the noisy output distributions of the messages. This is difficult to do because these output distributions are not iid. This is where \Cref{thm::bounds_uniform_type} is useful, as it allows us to use KL divergences over iid distributions in place of the KL divergence over the more complicated output distribution (since we can replace a hypergeometric distribution which undergoes noise with a multinomial distribution). Other obstacles include determining the covering number under KL divergence (see \Cref{sec::covering_def_results}).

\paragraph{Paper Organization}

We continue this section with the motivation and the notation. In \Cref{sec::covering}, we discuss how covering is used to determine the capacity of the noisy permutation channel along with some basics in covering. We prove \Cref{thm::bounds_uniform_type} and \Cref{thm::strict_positive} in \Cref{sec::main_divergence_results}.
 We prove all the parts of \Cref{thm::other_channels} in the appendix.

\subsection{Motivation}\label{sec:motiv}


The motivation for studying the permutation channel is that it captures a setting where codewords get reordered. This occurs in applications such as communication networks and biological storage systems. We briefly describe some of these applications. More details on these applications and other relevant work can be found in \cite{makur_2020}.

\paragraph{Communication Networks} Suppose we have a point-to-point communication network where the information is transmitted through a multipath routed network. Different packets are transmitted through different routes in the network, and each route has its own amount of latency, causing packets traveling on different routes to arrive at different times. The order in which the sender transmits packets is no longer preserved at the receiver end. Such a scenario is studied in \cite{walsh_multipath} where the authors are primarily concerned with reducing delay in their channel. Unlike our work, they do not consider noisy symbols. Another line of work which involves the permutation channel is on packet-switched networks. The errors explored in this work include insertions, deletions, and substitutions of symbols \cite{mladen_packetNetworks, mladen_perfectCodes}. Their work primarily focuses on building minimum distance codes and perfect codes for the permutation channel. 

\paragraph{DNA Storage Systems}

DNA-based storage systems are an attractive option for data storage due to its ability to withstand time and encode a very high-density of information \cite{yazdi_dnaTrends, erlich_dnaFountain}.   The state-of-the-art technology for storing information on DNA uses nucleotides with relatively small lengths (few hundreds) \cite{heckel_dnaStorage}. Each of these DNA molecules are stored in a pool without any regard to order. The different molecule types can be treated as symbols in the setting of the permutation channel. Noise in this channel models any error that can occur, whether it is in synthesizing the DNA molecules or in reading the molecules. DNA storage is also the motivation for studying the permutation channel in \cite{mladen_multisets, shomorony_noisyShuffling}.

\vspace{5mm}

As typical in information theory, a question of fundamental interest is to determine the capacity of channels. We determine the capacity of the noisy permutation channel in the strictly positive case, settling the problem introduced in \cite{makur_2020}. This setting differs from some of the models studied in the works described in the motivations, as it looks at the problem from a purely information theoretic standpoint and does not include assumptions which might be specific to the application. 

Among the works relevant to the motivations described, those that have some information theoretical flavors include \cite{mladen_multisets}, which deals with asymptotic bounds on rate, but for a fixed number of errors rather than probabilitistic errors. The work in \cite{heckel_dnaStorage} finds the capacity when the symbols are sampled randomly then read, something relevant to DNA models, but not to general permutation channels. The results in \cite{shomorony_noisyShuffling} are specifically for when the permuted objects are a string of symbols and the noisy process is applied to symbols on a string; the set of strings are permuted but symbols in each string are not.

Our results for when the DMC is the erasure channel are particularly interesting to DNA storage applications since the erased symbol can model deletion errors. Permutation channels with deletions are central to the work in \cite{mladen_multisets}, where the authors base their code constructions on Sidon sets and determine bounds on optimal codes for a given number of errors.  In our work, our errors are not fixed but probabilistic, and hence we find the probabilistic analogue of their bounds.

Also on the topic of deletions, one of the motivations for studying the noisy permutation channel in \cite{makur_2020} is the relation between permutation channels with erasures and the random deletion channel \cite{diggavi2001, mitzenmacher2006}. In \cite{mitzenmacher2006}, the author demonstrated a decoding scheme for the random deletion channel based on low density parity check (LDPC) codes. Their scheme can tolerate a reordering of the symbols, allowing it be a viable scheme for the permutation channel with erasures. However, their scheme requires the alphabet size to grow with the blocklength.

\subsection{Notation}\label{sec::notation}

The set of all probability distributions on $q$ symbols is defined as the probability simplex
\begin{align}
\Delta_{q-1} \eqdef \left\{(\pi_1,...,\pi_q): \sum_{i = 1}^q \pi_i = 1, 0 \leq\pi_i \leq 1\right\}\,.
\end{align}

For a $q \times \az$ DMC matrix  $P_{Z|X}$, we can express the individual transitions as
\begin{align}
P_{Z|X} = \begin{bmatrix}
p_{11} & p_{12} & \dots & p_{1k} \\
p_{21} & p_{22} & \dots & p_{2k} \\
\vdots & \vdots & \ddots & \vdots \\
p_{q1} & p_{q2} & \dots & p_{qk} 
\end{bmatrix}\,.
\end{align}
The values in each row of the matrix sums up to $1$ (i.e, the matrix is stochastic). Symbol $b \in \cX = [q]$ has probability $p_{bj}$ of becoming symbol $j \in \cY = [k]$. We can also write this probability as $P_{Z|X}(j|b)$. 
We say that the DMC matrix (or a submatrix) is \emph{strictly positive} if $p_{bj} > 0$ for all $b$ and $j$ in the matrix (or submatrix). 

For example, the DMC matrix for the BSC with crossover probability $\delta$, given in \eqref{eq::bsc_example_transition}, is written as
\begin{align}
P_{Z|X} = \begin{bmatrix}
1 - \delta & \delta \\
\delta & 1 - \delta
\end{bmatrix}\,.
\end{align}
If $0 < \delta < 1$, then this DMC matrix is strictly positive. 

Because of the uniform permutation step, the order of the symbols in $X^n$ does not matter. Thus, it is natural to consider the inputs to the channel as types classes rather than sequences. In light of this, we can describe the Markov chain of the noisy permutation channel as
\begin{align}\label{eq::channel_diagram2}
\pi \to X^n \to Z^n \to  Y^n
\end{align}
where each $\pi = (\pi_1,..,\pi_q) \in \Delta_{q-1} \cap \cP_n$ describes a type class. For communication, the sender has the freedom to encode messages into any $\pi$, and given $\pi$, the sequence $X^n$ can be any sequence in $T_n(\pi)$. The value of $\pi_b$ represents the proportion of positions in sequence $X^n$ which have symbol $b$. 

On the decoding end, the only relevant statistic the receiver would use from $Y^n$ is which type class $Y^n$ belongs to. Note that it is entirely equivalent to perform the permutation on the sequence $X^n$ first and then apply the DMC. In this case, we no longer need the random variable $Z^n$. Because of this, we also use $P_{Y|X}$ to specify the transition matrix, where $P_{Y|X}$ and $P_{Z|X}$ are the same and interchangeable.

Next, we specify a way to parameterize the distributions on $Y$.
We use the notation $Q_{Y|\mu}$ for $\mu = (\mu_1,...,\mu_k) \in \Delta_{\az - 1}$ to mean a distribution on symbols $\cY$ where the probability of symbol $j \in \cY$ is 
\begin{align}
Q_{Y| \mu} (j) = \mu_j\,.
\end{align}
The distribution $Q_{Y|\mu}^n$ is the multinomial distribution with parameters $\mu$ and number of independent trials $n$.
These distributions do not (directly) relate the permutation channel; we define them since they are important for our analysis.

On the other hand, the distribution $P_{Y^n|\pi}$ refers the the distribution on sequences $Y^n$ when $\pi \in \cP_n$ is the input to the noisy permutation channel on $n$ letters as described in \eqref{eq::channel_diagram2}. Note that in general $P_{Y^n|\pi}$ is \emph{not} a multinomial distribution. As seen in \Cref{thm::bounds_uniform_type},
\begin{align}\label{eq::composition_types}
P_{Y^n|\pi} = P_{Y|X}^n \circ U   
\end{align}
where $U$ is a uniform distribution on $T_n(\pi)$. Both represent the distribution on the output of the noisy permutation channel. Permuting the input symbols gives a sequence in the support of $U$, and then each permuted symbol goes through the transition probabilities $P_{Y|X}$ independently.

When it is clear what $\pi$ is, we use $P_Y$ to mean the marginal distribution for each 
$Y_t$ in the sequence $Y^n \sim P_{Y|X}^n \circ U $. This distribution does not depend on the index $t$ since $U$ is uniform on all permutations.

Throughout this work, we use $\log$ to mean the natural logarithm.





\section{Covering Converse}\label{sec::covering}

Our core method for finding our new results is to use KL divergence covering of the probability simplex. We first show how covering can be applied to the noisy permutation channel and then give the necessary covering results. 

\subsection{Covering Basics}

The main concept of our proof uses covering ideas similar to \cite[Theorem 1]{yang_barron} in order  to upper bound the mutual information $I(\pi; Y^n)$. In summary, we need to find a set of covering centers which are close in \emph{Kullback-Leibler (KL) divergence} to all the possible distributions on $Y^n$ that can occur as outputs of the noisy permutation channel. Our set of centers need not be a possible distribution over $Y^n$ generated by the channel. We choose to use multinomial distributions as our set of covering centers. 

Let $\mathcal{N}_n$ be a discrete set in $\Delta_{\az-1}$ which we specify (later) for each $n$ (this will be the covering centers). Mutual information has the property that
\begin{align} \label{eq::MI_first}
I(\pi; Y^n) \leq \max_{\pi} D(P_{Y^n|\pi}  \| \tilde Q_{Y^n})\,.
\end{align}

The above holds for any $\tilde Q_{Y^n}$, thus we can choose
\begin{align}
\tilde Q_{Y^n}(y^n) &= \frac{1}{|\mathcal{N}_n|}\sum_{ \mu \in \mathcal{N}_n} Q^n_{Y|\mu}(y^n) 
\\&= \frac{1}{|\mathcal{N}_n|}\sum_{\mu \in \mathcal{N}_n}  \prod_{t = 1}^n Q_{Y|\mu} (y_t)\,.
\end{align}


The following proposition is the main tool of all our converse results.

\begin{fproposition}[Covering for Noisy Permutation Channels]\label{prop::covering_argument}
Suppose that for the noisy permutation channel with DMC $P_{Y|X}$, we have that for any $\pi \in \cP_n$, 
\begin{align}\label{eq::poisson_covering}
D(P_{Y|X}^n \circ U  \| Q^n_{Y}) \leq nD(P_Y \| Q_{Y}) + f(n)
\end{align}
where $U$ is uniform on the type $T_{n}(\pi)$, $P_Y$ is the marginal distribution of $P_{Y|X}^n \circ U$ and $f$ is only a function of $n$ and $P_{Y|X}$. Then 
\begin{align}
\permcap(P_{Y|X}) \leq \frac{\rank(P_{Y|X}) - 1}{2} + \lim_{n \to \infty}\frac{f(n)}{\log n}\,.
\end{align}
\end{fproposition}

In \Cref{prop::covering_argument}, when the DMC is strictly positive, the $f(n)$ term is constant in $n$ (which is shown via \Cref{thm::bounds_uniform_type} and gives the proof for \Cref{thm::strict_positive}). However, when the DMC is not strictly positive, $f(n)$ is not necessarily constant in $n$. Non-constant values of $f(n)$ are used in deriving some of the results in \Cref{thm::other_channels}.

For the proof, we need to define for any $\pi \in \Delta_{q - 1}$
\begin{align}
\mu^M(\pi) \eqdef \left(\sum_{i} \pi_i p_{i1}, ..., \sum_{i} \pi_i p_{ik}\right)\,.
\end{align}
The vector $\mu^M(\pi)$ is the mean (we use `M' as short for mean) of the distribution $P_{Y^n|\pi} = P_{Y|X}^n \circ U$. Note that $P_Y(j) = \sum_{i} \pi_i p_{ij}$. Also if $\mu = \mu^M(\pi)$, we can write $P_Y = Q_{Y|\mu}$.

\begin{proof}
Following techniques used in the proof of \cite[Theorem 1]{yang_barron}, we can upper bound the mutual information given in \eqref{eq::MI_first} by
\begin{align} \label{eq::mi_2}
I(\pi; Y^n) \leq \log | \mathcal{N}_n | + \max_{\pi \in \cP_n}\min_{\bar \mu \in \mathcal{N}_n}D(P_{Y^n|\pi}  \| Q^n_{Y|\bar\mu})\,.
\end{align}

To specify $\mathcal{N}_n$, first define
\begin{align}
\mathcal{L}(P_{Y|X}) = \bigcup_{\pi \in \Delta_{\az-1}} \mu^{M}(\pi)\,.
\end{align}
This is the space of all possible marginals $P_Y$.

Let $\mathcal{N}_n$ be a covering of $\mathcal{L}(P_{Y|X})$ under KL divergence with covering radius $1/n$. In other words, $\mathcal{N}_n = \{\bar\mu^{(1)},...,\bar \mu^{(m)} \}$ so that 
\begin{align}
 \max_{\mu \in \mathcal{L}(P_{Y|X})}\min_{\bar \mu \in \mathcal{N}_n} D(Q_{Y|\mu} \| Q_{Y|\bar \mu}) \leq \frac{1}{n}\,.
\end{align}

Let $\ell$ be the dimension of $\mathcal{L}(P_{Z|X})$. Using divergence covering results and the result specifically about covering an $\ell$-dimensional subspace (see next part \Cref{sec::covering_def_results}),
\begin{align}
|\mathcal{N}_n| &\leq  C(q, \ell) \left(\frac{\ell}{1/n}\right)^{\frac{\ell}{2}} 
\label{eq::covering_bound}
\end{align}
where $C(q, \ell)$ depends on $q$ and $\ell$ but not on $n$.

Starting with \eqref{eq::mi_2} and substituting in assumption \eqref{eq::poisson_covering} gives
\begin{align}
I(\pi; Y^n) 
&\leq \log  \left(C(q, \ell)\left(\frac{\ell}{1/n}\right)^{\frac{\ell}{2}}\right)  \eqlinebreakshort+ f(n) + \max_{\pi \in \cP_n}\min_{\bar\mu \in \mathcal{N}_n}nD(P_Y \|Q_{Y|\bar \mu}) \\
&\leq \frac{\ell}{2} \log n + \log C(q, \ell) + \frac{\ell}{2} \log  \ell + f(n) +  n \frac{1}{n}\\
&\leq \frac{\ell}{2} \log n + c' + f(n)
\end{align}
where $c'$ is a constant which does not depend on $n$. 

For the noisy permutation channel, recall that the rate is defined as \eqref{eq::rate_def}. 
Since asymptotically $\log M \leq I(\pi, Y^n) \leq \frac{\ell}{2} \log n + c' + f(n)$, we have
\begin{align}
R \leq \frac{\ell}{2} + \frac{c'}{\log n} + \frac{f(n)}{\log n} \to  \frac{\ell}{2} + \lim_{n \to \infty} \frac{f(n)}{\log n}\,.\label{eq::R_compute_ell}
\end{align}

It remains to compute $\ell$. Let $r = \rank(P_{Z|X})$. When the domain is any vector in $\bbR^q$, the image space of this (left) vector multiplied by $P_{Z|X}$ is $r$ dimensional. But since we are restricting the domain and image to probability vectors, this adds an additional constraint to the image space and reduces the dimension by $1$, giving that $\ell = \rank(P_{Z|X}) - 1$. Substituting this into \eqref{eq::R_compute_ell} gives an upper bound for the capacity of the noisy permutation channel.

\end{proof}

\subsection{Covering Definition and Results}
\label{sec::covering_def_results}
In order to show \eqref{eq::covering_bound}, we have the following definition and results. 
A \emph{KL divergence covering} is a set of centers in $\Delta_{\az -1}$ so that every point in $\Delta_{\az -1}$ is within some KL distance of one of the centers. Let $\varepsilon$ be this distance.  Since KL divergence is not symmetric, we specify that KL distance is computed where the covering center is placed in the second argument of the KL divergence. This is made explicit in the following definition of a \emph{covering number}.
\begin{fdefinition}[Divergence Covering Number]\label{def::covering}
\begin{align}
M(\az, \varepsilon) &= \inf\{m: \exists\{Q_1,...,Q_m\}  \eqlinebreakshort\text{ s.t } \max_{P \in \Delta_{\az-1}}\min_{Q_i} D(P||Q_i) \leq \varepsilon\}\,.
\end{align}

Let $M(\az, \varepsilon, \cB)$ be defined like $M(\az, \varepsilon)$ except that $P \in \cB$ for a subset $\cB \subset \Delta_{\az-1}$. 
\end{fdefinition}

We need upper bounds on the KL divergence covering number in order to get converse results for the permutation channel. One such upper bound is the following:

\begin{ftheorem}[Upper Bound on Divergence Covering]\label{thm::covering_non_random}

For $0 < \varepsilon \leq 1$, 
\begin{align}
M(\az, \varepsilon) \leq c^{\az-1}\left(\frac{\az-1}{\varepsilon}\right)^{\frac{\az-1}{2}}
\end{align}
for some constant $c$.
\end{ftheorem}

The above result is sufficient for showing our theorems. However, tighter bounds do exist (see \cite{phdthesis}). In addition to an upper bound on the KL divergence covering, we also need an additional result which allows us to use covering numbers over the whole simplex to get covering numbers over certain subsets of the simplex.

\begin{fproposition}\label{prop::cover_subset}
For $\cB \subset \Delta_{\az-1}$, suppose there is a stochastic matrix $F$ which maps $\Delta_{q - 1}$ onto $\cB$. Suppose that $\cB$ is a space of dimension $\ell - 1$ (or likewise, $F$ has rank $\ell$). Then,
\begin{align}
M(\az, \varepsilon, \cB) \leq \binom{q}{\ell}M(\ell, \varepsilon)\,.
\end{align}
\end{fproposition}
The proofs are in \Cref{sec::covering_results}. More discussion on KL divergence covering can be found in \cite{phdthesis}.

\section{Divergence Under Fixed Types}\label{sec::main_divergence_results}

For computing our converse bounds, we need to determine the expression \eqref{eq::poisson_covering} for our DMC matrices. This is where we need \Cref{thm::bounds_uniform_type} which gives the divergence between noisy observations of a fixed type compared to an iid distribution.


We prove \Cref{thm::bounds_uniform_type} by first showing some relevant intermediate results. The techniques in these intermediate results are also useful for when $P_{Z|X}$ is not strictly positive and therefore relevant for showing some of the results in  \Cref{thm::other_channels}. Before doing so, we briefly discuss the constant of \Cref{thm::bounds_uniform_type} and how it is tight. 

\subsection{Constant of \Cref{thm::bounds_uniform_type}}

Here we show that the constant $c$ in \Cref{thm::bounds_uniform_type} is sharp (cannot be improved to $o(1)$). One tool we need is the following theorem by Marton \cite{martonBlowUp}:
\begin{lemma}[Marton's Transportation Inequality]
\label{lem::marton_info_coupling}
Let $X^n \sim \prod_{t = 1}^n P_{X_t}$ and $\hat X^n \sim P_{\hat X^n}$. Then there exists a joint probability measure $P_{X^n, \hat X^n}$ with these given marginals such that
\begin{align}
\frac{1}{n} \bbE[d(X^n, \hat X^n)] &= \frac{1}{n} \sum_{t = 1}^n \bbP[X_t \neq \hat X_t] \\
& \leq \left(\frac{1}{n} D\left(P_{\hat X^n} \bigg\| \prod_{t = 1}^n P_{X_t}\right) \right)^{1/2} 
\end{align}
where $d(X, Y)$ is the Hamming distance. 
\end{lemma}

Suppose that we are only working with $2$ symbols, $\{1, 2\}$, for the space of $X$ and $Y$. Using the notation in \Cref{thm::bounds_uniform_type}, let $Y^n \sim P_{Y|X}^n \circ U$ and $\hat Y^n \sim Q_{Y}^n$, where we set $Q_Y = P_Y$. Choose $P = (1/2, 1/2)$ and $P_{Y|X}$ to be that of a BSC with crossover probability $\delta$. This gives that distribution $P_Y$ is uniform on the two symbols. 

We choose $\delta = 1/n$, which is non-zero but small enough so that in expectation, $X^n$ and $Y^n$, as well as  $\hat X^n$ and $\hat Y^n$, will differ by $1$. Define function $\#1(\cdot)$ to be mean the number of $1$'s in a sequence. Then
\begin{align}
    \bbE |\#1(Y^n) - n/2 |  &=  \bbE |\#1(Y^n) - \#1(X^n)|  
    \\& \leq \bbE[d(Y^n, X^n)] 
    \\& = 1
     \\\bbE |\#1(\hat Y^n) - \#1(\hat X^n) | &\leq \bbE[d(\hat Y^n, \hat X^n)] 
     \\&= 1 
\end{align}

The number of $1$'s in $\hat X^n$ will differ from its mean by roughly the standard deviation. To be precise, applying a result from \cite{Berend2013}, we have that
\begin{align}
    \bbE |\#1(\hat X^n) - n/2 | \geq \frac{1}{2\sqrt{2}}\sqrt{n}\,.
\end{align}
Using the above and multiple applications of the triangle inequality, we can compute that no matter the coupling chosen
\begin{align}
    \bbE[d(Y^n, \hat Y^n)] &\geq \bbE |\#1(\hat Y^n) - \#1(Y^n)| 
    \\ & \geq \bbE |\#1(\hat Y^n) - n/2 | - \bbE |\#1( Y^n) - n/2 |
    \\ & \geq \bbE |\#1(\hat X^n) - n/2 |  - \bbE |\#1(\hat Y^n) - \#1(\hat X^n) | 
    \eqlinebreak
    - \bbE |\#1( Y^n) - n/2 |
    \\&= \frac{1}{2\sqrt{2}}\sqrt{n} - 2
\end{align}
We can assume that $n$ is large, so that the Hamming distance can be more simply lower bounded by
\begin{align}
    \bbE[d(Y^n, \hat Y^n)] \geq \frac{1}{4} \sqrt{n}\,.
\end{align}

Combining this Hamming distance with \Cref{lem::marton_info_coupling} 
gets a lower bound 
\begin{align}
\frac{1}{4\sqrt{n}} &\leq \frac{1}{n} \bbE[d(Y^n, \hat Y^n)] \\
&\leq  \left(\frac{1}{n}  D(P_{Y|X}^n \circ U \| Q_Y^n)\right)^{1/2} \\
&\leq  \left(\frac{1}{n} (n D(P_Y\|P_Y) + c)\right)^{1/2} \\
&= \frac{\sqrt{c}}{\sqrt{n}}\,.
\end{align}
Improving $c$ to be $o(1)$ in $n$ would violate this lower bound. 


Intuitively, consider what happens when we let $P_{Y|X}$ be the identity matrix where $Y = X$. In such a case, \Cref{thm::bounds_uniform_type} is not true (in order to get a true statement, the constant $c$ should be replaced with a value that grows logarithmically with $n$, see \Cref{sec::no_strictly_positive}). This is the same setting as the example given above but with $\delta = 0$. It is clear here that $\hat Y^n$ likely has $\sqrt{n}$ deviations in the number of $1$'s from the mean whereas any sequence $Y^n$ has exactly $n/2$ number of $1$'s. This creates an expected Hamming distance of $\sqrt{n}$. Slightly increasing $\delta$ above zero will not change the Hamming distance by much but will make $P_{Z|X}$ strictly positive.

\subsection{Expression for Divergence Under Fixed Types}

In order to show \Cref{thm::bounds_uniform_type}, we need some intermediary results about how to work with the quantity $D( P_{Y|X}^n \circ U \| Q_Y^n)$. The next proposition does this and can be used for any $P_{Y|X}$, not just those which are strictly positive. 

\begin{fproposition}\label{prop::compute_divergence}

Let $U$ be uniform on the type $T_n(P)$ and $(X,Y)^n$ be iid from $(P \times P_{Y|X})$. Let $P_Y$ be the marginal distribution of $Y$ under $(P \times P_{Y|X})$. Then for all $Q_Y$, 
\begin{align}
D&( P_{Y|X}^n \circ U \| Q_Y^n) =nD(P_{Y} \| Q_Y) 
\\ &
+ \sum_{y^n \in \cY^n} \bbP[Y^n = y^n | A = 1] \log \frac{\bbP[ A = 1 | Y^n = y^n]}{\bbP[A = 1]}\label{eq::divergence_diff_bound}
\end{align}	
where $A= \bbI\{X^n \in T_n(P)\}$ and under $\bbP$ the sequence $(X,Y)^n$ is iid from $(P \times P_{Y|X})$.
\end{fproposition}

The second term on the right-hand side of \eqref{eq::divergence_diff_bound} can be written as an expected value:
\begin{align}
\sum_{y^n \in \cY^n} & \bbP[Y^n = y^n | A = 1] \log \frac{\bbP[ A = 1 | Y^n = y^n]}{\bbP[A = 1]} 
\\ & = \bbE_{(X, Y)^n \sim (P \times P_{Y|X})}
\eqlinebreakshort 
\Bigg[\log \frac{\bbP_{(\tilde X,\tilde Y)^n \sim (P \times P_{Y|X})}[\tilde X^n \in T(P)|\tilde Y^n = Y^n]}{\bbP_{(\tilde X, \tilde Y)^n \sim (P \times P_{Y|X})}[\tilde X^n \in T(P)]}
\\ & \quad \quad \quad
\bigg| (X,Y)^n 
\text{ where } X^n \in T(P)\Bigg]
\end{align}

For ease of notation, we choose to express the term above as
\begin{align}
\bbE\left[\log \frac{ \bbP[\tilde A=1|\tilde Y^n = Y^n]}{ \bbP[\tilde A=1]} \bigg| A=1\right]\label{eq::expected_additional_term}
\end{align}
where the $\tilde \cdot$  notation emphasizes that the variables are associated with an independent copy $(\tilde X, \tilde Y)^n$ drawn from the same distribution $(P \times P_{Y|X})$ as $(X, Y)^n$ and where $\tilde A= \bbI\{\tilde X^n \in T_n(P)\}$.


\begin{proof}

Note that $(P_{Y|X}^n \circ U) (y^n) = \bbP[Y^n = y^n | A = 1]$.

\begin{align}
D \eqstartshort ( P_{Y|X}^n \circ U \| Q_Y^n) \eqstartnotshort = \sum_{y^n} \bbP[Y^n = y^n | A = 1] \eqlinebreakshort \log \frac{\bbP[Y^n = y^n | A = 1]}{Q_Y^n(y^n)}\\
& = \sum_{y^n} \bbP[Y^n = y^n | A = 1] \eqlinebreakshort\log \frac{\bbP[ A = 1 | Y^n = y^n] \bbP[Y^n = y^n]}{\bbP[A = 1]Q_Y^n(y^n)}\\
&=  \bbE\left[ \log \frac{P_{Y}^n(Y^n)}{Q_Y^n(Y^n)}  \bigg| A = 1 \right]  \eqlinebreakshort+ \sum_{y^n} \bbP[Y^n = y^n | A = 1] \log \frac{\bbP[ A = 1 | Y^n = y^n]}{\bbP[A = 1]}\\
&=  \bbE\left[ \log \frac{P_{Y}^n(Y^n)}{Q_Y^n(Y^n)}  \bigg| A = 1 \right]  \eqlinebreakshort+ \bbE\left[\log \frac{\bbP[\tilde A = 1|\tilde Y^n = Y^n]}{\bbP[\tilde A = 1]} \bigg| A = 1 \right]\,.
\end{align}

The marginal distribution $P_Y(a)$ is also the probability that any position $t$ in sequence $Y^n$ takes the value $a$, i.e. $P_Y(a) = \bbP[Y_t = a | A = 1]$. This occurs since $U$ is uniform on all permutations of type $T_n(P)$. 
We get for the first term in the sum, 
\begin{align}
\bbE \eqstartshort \left[ \log \frac{P_{Y}^n(Y^n)}{Q_Y^n(Y^n)}  \bigg| A = 1 \right]
\\
&=\sum_{y^n} \bbP[Y^n = y^n | A = 1]\log \frac{P_{Y}^n(y^n)}{Q^n_Y(y^n)}\\
&= \sum_{y^n} \bbP[Y^n = y^n | A = 1] \sum_{a} n \frac{|\{t:y_t = a\} |}{n}\log \frac{P_{Y}(a)}{Q_Y(a)}\\
& = n \sum_{a} P_Y(a) \log \frac{P_{Y}(a)}{Q_Y(a)}\\
& = n D(P_Y \| Q_Y)\,.
\end{align}
This gives the result \eqref{eq::divergence_diff_bound}.
\end{proof}

We can separate \eqref{eq::expected_additional_term} into two additive terms due to the logarithm. The next lemma can be used to compute one of these terms.

\begin{flemma}\label{lem::prob_a}
Let $P  = (p_1,...,p_q) \in \cP_n$ and let $A= \bbI\{X^n \in T_n(P)\}$. If $(X, Y)^n$ is drawn iid from $(P \times P_{Y|X})$, then
\begin{align}
\log \frac{1}{\bbP[A = 1]} &\leq -\frac{1}{2} \log n + \sum_{i: p_i > 0} \frac{1}{2} \log p_i n  \eqlinebreakshort+ \frac{q-1}{2} \log 2 \pi + \frac{1}{12 n}\,. \label{eq::solve_prob_a}
\end{align}
\end{flemma}

For this proof, we use a Stirling approximation type bound from \cite{robbins_stiring}: For positive integers $n$, 
\begin{align}
\sqrt{2 \pi n} \left(\frac{n}{e}\right)^{n} e^{\frac{1}{12n + 1}} \leq n! \leq \sqrt{2 \pi n} \left(\frac{n}{e}\right)^{n} e^{\frac{1}{12n}}\label{eq::stirling_12}\,.
\end{align}

\begin{proof}
We assume that all $p_i > 0$ since we can always reduce $P$ to a shorter vector and decrease $q$. 

The probability that a specific type occurs is given by the multinomial distribution. 
\begin{align}
-\eqstartshort \log \bbP[A = 1] \eqbreakshort
& = -\log \left(\frac{n!}{\prod_{i = 1}^q(p_i n)!} \prod_{i = 1}^q p_i^{p_i n}\right)\\
& = -\log \left(\frac{n!}{n^n} \right) - \log \left(\prod_{i = 1}^q  \frac{(p_i n)^{p_i n}}{(p_i n)!} \right) \\
& \leq n - \frac{1}{2} \log n - \frac{1}{2} \log 2\pi \eqlinebreakshort + \sum_{i = 1}^{q} \left( -p_i n + \frac{1}{2} \log p_i n + \frac{1}{2} \log 2 \pi + \frac{q}{12 n} \right)\,.
\end{align}
We used \eqref{eq::stirling_12} in the last inequality (we can do this since each $p_i n$ is an integer greater than $0$). Combining terms gives the result. 
\end{proof}

\subsubsection{\Cref{thm::bounds_uniform_type} Without the Strictly Positive Requirement}\label{sec::no_strictly_positive}

To illustrate how to use \Cref{prop::compute_divergence} and \Cref{lem::prob_a}, we compute an upper bound with the same form as the upper bound given in \eqref{eq::bounds_uniform_type_eq} of \Cref{thm::bounds_uniform_type} for all $P_{Y|X}$. We briefly mentioned above that if we remove the strictly positive requirement for $P_{Y|X}$ in \Cref{thm::bounds_uniform_type}, in the worst case, the constant $c$ would need to be replaced with a logarithmic term. To be exact, if $P_{Y|X}$ is not strictly positive, $c$ needs to be replaced with a poly-logarithmic term in $n$. Using \Cref{prop::compute_divergence} and \Cref{lem::prob_a}, we can get an upper bound on $c$ with
\begin{align}
c &= \bbE\left[\log \frac{\bbP[\tilde A = 1|\tilde Y^n = Y^n]}{\bbP[\tilde A = 1]} \bigg| A = 1 \right]\\
& \leq  \frac{q-1}{2}\log  n + c' + \bbE\left[\log \bbP[\tilde A = 1|\tilde Y^n = Y^n] \bigg| A = 1 \right]\\
& \leq  \frac{q-1}{2}\log  n + c'\,.
\end{align}
We used the fact that the largest value $\bbP[\tilde A = 1|\tilde Y^n = Y^n]$ can take is $1$ since it is a probability. 
The inequality is tight when $P_{Y|X}$ is the identity matrix (when $Y=X$). One example when $Y=X$ is the BSC with no noise ($\delta = 0$) example we stated earlier, where \eqref{eq::bsc_no_noise_KL} holds. 

\subsection{Concentration of Sums of Independent Variables}
\label{sec::petrov_concentration}

To show \Cref{thm::bounds_uniform_type}, we need to compute \eqref{eq::expected_additional_term}. We need to determine the probability of $\tilde A = 1$, which is the event that $\tilde X^n$ has a particular type, under certain conditions. Showing that $\tilde X^n$ has a particular type can be equated to the problem of randomly throwing balls into some set of bins and looking at the number of balls which fall into each bin. To help us bound the probability a certain number of balls falls into a particular bin, we make use of the following:

The concentration function $Q(Z; \lambda)$ of random variable $Z$ is defined by 
\begin{align}
    Q(Z; \lambda) = \sup_{z} \bbP[z \leq Z \leq z + \lambda]
\end{align}
for every $\lambda \geq 0$ \cite{petrov_sums}. 
Let $S_n = \sum_{i = 1}^n W_i$ where $W_i$ are independent random variables.

\begin{theorem}[Petrov \cite{petrov_sums}]\label{thm::petrov_sum}
Let the numbers $a_i$ and $b_i$ be such that 
\begin{align}
    \bbP\left[W_i - a_i \leq - \frac{\lambda_i}{2} \right] &\geq b_i \\
    \bbP\left[W_i - a_i \geq \frac{\lambda_i}{2} \right] &\geq b_i 
\end{align}
for $i = 1,...,n$. Then there exists a universal constant $\alpha$ so that 
\begin{align}
    Q(S_n; \lambda) \leq \alpha \lambda \left(\sum_{i = 1}^n \lambda_i^2 b_i \right)^{-1/2}
\end{align}
for every positive $\lambda_1,...,\lambda_n$ none of which exceeds $\lambda$.
\end{theorem}
To apply \Cref{thm::petrov_sum} to our problem, each $W_i$ is a Bernoulli random variable where the probability that $W_i = 1$ is $p_i$. Let $b_i = \min\{p_i, 1-p_i\}$. We can fix $a_i = 1/2$. We can also fix $\lambda_i = 1/2$ and $\lambda = 1/2$, though this exact value does not matter so long as $\lambda < 1 - \varepsilon$ for a small $\varepsilon > 0$. 

This gives that for any integer $z$ 
\begin{align}
    \bbP[S_n = z] &\leq Q(S_n; 1/2)  \\
    &\leq \frac{\alpha (1/2)}{\sqrt{\sum_{i = 1}^n (1/2)^2 \min \{p_i, 1-p_i \} }}\\
    & \leq \frac{\alpha}{\sqrt{\sum_{i = 1}^n \min \{p_i, 1-p_i \} }}\,. \label{eq::variance_proxy}
\end{align}
We use this in the next lemma which is the key to computing the second term in \eqref{eq::divergence_diff_bound}.

\begin{flemma}\label{lem::balls_in_bins}

Suppose there are $n$ balls which are thrown into one of $q$ bins. Each ball is thrown independently, and for the $i$-th ball, the probability of landing in bin $b$ is $p_{i,b}$. 

Let $N_b$ be the ball count of the $b$-th bin.
Then if $\pi_b > 0$ for all $b$ and $\sum_{b} \pi_b = 1$, we have
\begin{align}
\bbP [N_1=n \pi_1,\dots, N_q = n \pi_q] \leq \frac{\alpha^{q-1}}{n^{(q-1)/2}  \sqrt{B}}
\end{align}
where
\begin{align}
B &= c_{*}^{q-1}\frac{\prod_b \pi_b }{\pi_{max}}\\
c_{*} &= \min_{i} \frac{c_{-}(i)}{c_{+}(i)}\\
c_{-}(i) &= \min_{b} \frac{p_{i,b}}{\pi_b}\\
c_{+}(i) &= \max_{b} \frac{p_{i,b}}{\pi_b}\\
\pi_{max} &= \max_{b} \pi_b
\end{align}
and $\alpha$ is the universal constant used in \Cref{thm::petrov_sum}. 
\end{flemma}

\begin{proof}
For notation, let $W_{i, b}$ be the indicator variable of whether ball $i$ was thrown into bin $b$. We can express $N_b = \sum_{i = 1}^n W_{i, b}$.
Arrange the indices so that $\pi_1 \leq \pi_2 \dots \leq \pi_q$.

First observe that
\begin{align} 
\bbP \eqstartshort [N_1=n \pi_1,\dots, N_q = n \pi_q]  
\\&= \prod_{b = 1}^q \bbP [N_b = n \pi_b |N_1=n \pi_1,\dots, N_{b-1} = n \pi_{b-1}]\,.\label{eq::product_of_conditionals}
\end{align}
For $b = q$, 
\begin{align}
\bbP [N_b = n \pi_b |N_1=n \pi_1,\dots, N_{b-1} = n \pi_{b-1}] = 1\,.
\end{align}
%
For $b < q$, we can compute for any $i$ that
\begin{align}
 \min \eqstartshort \left\{\frac{p_{i,b}}{\sum_{a = b}^q p_{i, a}}, 1 -\frac{p_{i,b}}{\sum_{a = b}^q p_{i, a}}\right\} \eqbreakshort
& =  \min\left\{\frac{p_{i,b}}{\sum_{a = b}^q p_{i, a}}, \frac{\sum_{a > b}^q p_{i,a}}{\sum_{a = b}^q p_{i, a}}\right\}  \\
& = \min\left\{ \frac{\pi_b\frac{p_{i,b}}{\pi_b}}{\sum_{a = b}^q \pi_a\frac{p_{i, a}}{\pi_a}}, \frac{ \sum_{a > b}^q \pi_a \frac{ p_{i,a}}{\pi_a}}{\sum_{a = b}^q \pi_a\frac{p_{i, a}}{\pi_a}}\right\}\\
& \geq \frac{\min_{a} \frac{p_{i,a}}{\pi_a}}{\max_{a} \frac{p_{i,a}}{\pi_a}} \min\left\{ \frac{\pi_b}{\sum_{a = b}^q \pi_a}, \frac{\sum_{a > b}^q \pi_a}{\sum_{a = b}^q \pi_a}\right\}\\
& \geq \min_{i} \frac{c_{-}(i)}{c_{+}(i)} \frac{1}{\sum_{a = b}^q \pi_a} \min\left\{\pi_b, \sum_{a > b}^q \pi_a \right\}\\
& = c_{*} \frac{\pi_b}{\sum_{a = b}^q \pi_a} \,.
\end{align}
We get the last equality because we have arranged $\pi_b$ in increasing order.
Hence by \eqref{eq::variance_proxy}
\begin{align}
\bbP\eqstartshort[N_b = n \pi_b | N_1 = n \pi_1,\dots, N_{b-1} = n \pi_{b-1}] 
\eqbreakshort
&\leq \frac{\alpha}{\sqrt{\left(n - \sum_{a = 1}^{b-1} n \pi_a\right) c_{*} \frac{\pi_b}{\sum_{a = b}^q \pi_a}  }} \\ 
&= \frac{\alpha}{\sqrt{\frac{n - \sum_{a = 1}^{b-1} n \pi_a}{n} nc_{*}\frac{\pi_b}{\sum_{a = b}^q \pi_a}  }} \\ 
& = \frac{\alpha}{n^{1/2}\sqrt{c_{*} {\pi_b} }}
\end{align}
where we used that $n - \sum_{a = 1}^{b-1} n \pi_a  =  n \sum_{a = b}^q \pi_a$ to get the last inequality. 
Taking a product of all terms in \eqref{eq::product_of_conditionals}, gives
\begin{align}
\bbP \eqstartshort [N_1=n \pi_1,\dots, N_q = n \pi_q] \eqbreakshort
&\leq \prod_{b = 1}^{q-1} \frac{\alpha}{n^{1/2}\sqrt{ c_{*} {\pi_b} }}\\
&= \frac{\alpha^{q-1}}{n^{(q-1)/2} \sqrt{ c_{*} ^{q-1}\prod_{b = 1}^{q-1} \pi_b }} \\
&= \frac{\alpha^{q-1}}{n^{(q-1)/2} \sqrt{ c_{*}^{q-1}\frac{\prod_b \pi_b }{\pi_{q}}}} \\
&= \frac{\alpha^{q-1}}{n^{(q-1)/2} \sqrt{B}}\,.
\end{align}
\end{proof}

\subsection{Completing Proof of \Cref{thm::bounds_uniform_type} and Determining Capacity.}

We can now use \Cref{lem::balls_in_bins} to prove \Cref{thm::bounds_uniform_type}.

\begin{proof}[Proof of \Cref{thm::bounds_uniform_type}]
We show the lower bound first, which is easier to show. Using \Cref{prop::compute_divergence}, we need only to show that 
\begin{align}
\bbE\left[\log \frac{\bbP[\tilde A = 1|\tilde Y^n = Y^n]}{\bbP[\tilde A=1]} \bigg| A=1\right] \geq 0\,.
\end{align} 
We do this by
\begin{align}
\bbE \eqstartshort \left[\log \frac{\bbP[\tilde A = 1|\tilde Y^n = Y^n]}{\bbP[\tilde A=1]} \bigg| A=1\right]  \eqbreakshort
& = \sum_{y^n} \bbP[Y^n = y^n | A = 1] \log \frac{\bbP[A=1|Y^n = y^n]}{\bbP[A=1]} \\
& = \sum_{y^n} \bbP[Y^n = y^n | A = 1] \eqlinebreakshort\log \frac{\bbP[Y^n = y^n|A = 1] \bbP[A=1]}{\bbP[Y^n = y^n]\bbP[A=1]} \\
& = \sum_{y^n} \bbP[Y^n = y^n | A = 1] \log \frac{\bbP[Y^n = y^n|A = 1] }{\bbP[Y^n = y^n]}\\
& = D( \bbP[Y^n | A = 1] \| \bbP[Y^n])\\
& \geq 0
\end{align}
since divergences are always non-negative. 

To get the upper bound, we use
\begin{align}
\bbE \eqstartshort \left[\log \frac{\bbP[\tilde A = 1|\tilde Y^n = Y^n]}{\bbP[A=1]} \bigg| A=1\right] 
\\&= \bbE\left[\log \bbP[\tilde A = 1|\tilde Y^n = Y^n] \bigg| \tilde A=1\right] - \log \bbP[A=1] \label{eq::constant_term_to_compute}
\end{align}
and use \Cref{lem::balls_in_bins} for the first term in the sum and \Cref{lem::prob_a} for the second term in the sum.

\Cref{lem::balls_in_bins} applies to the first term because, given some $Y^n$, finding the probability that the type of $X^n$ is in $T_n(P)$ is equivalent to finding the number of balls which are randomly thrown into each bin. We want to determine the probability that $X^n$ is in $T_n(P)$ when $(X, Y)^n \sim (P_{Y|X} \times P)$. 

Let $P$ of $T_n(P)$ be expressed as $P = (\pi_1 , ..., \pi_q )\in \cP_n$. This implies that $\pi_b = \bbP[X = b]$. Let the balls described in \Cref{lem::balls_in_bins} be each of the elements of $Y^n$. If $Y_i = y_i$, then let $p_{i,b} = \bbP[X_i = b| Y_i = y_i] = \bbP[X = b| Y = y_i]$ (because the symbols are iid). This way $p_{i,b}$ is appropriately the probability that the $i$th symbol lands in bin $b$. As in \Cref{lem::balls_in_bins}, $N_b$ is the number of balls in bin $b$. Then the probability that $X^n \in T_n(P)$ is equivalent to 
$\bbP[N_1 = n \pi_1,\dots, N_q = n \pi_q]$. This is computed for a specific value of $Y^n$, but notice that the expression we derived for $\bbP[N_1 = n \pi_1,\dots, N_q = n \pi_q]$ in \Cref{lem::balls_in_bins} does not depend on $Y^n$. 

Before computing the rest of the expression, we need to pay particular attention to the case when there exists a $b$ such that $\pi_b = 0$. If $\pi_b =0 $, then $\bbP[X = b] = 0$, which would also imply that $p_{i,b} = 0$ for all $i$. In this case, we can remove the symbol $b$ (or bin $b$ in the interpretation of \Cref{lem::balls_in_bins}) from consideration and apply \Cref{lem::balls_in_bins} to just the symbols with non-zero probability. We can always reorder the symbols, so that the first $q'$ of the $q$ symbols all have $\pi_b > 0$ and the remaining $b >q'$ are such that $\pi_b = 0$. Like in \Cref{lem::balls_in_bins}, we can define
\begin{align}
B &=  c_{*}^{q'-1}\frac{\prod_{b = 1}^{q'} \pi_b }{\pi_{max}}\\
c_{*} &= \min_{i} \frac{c_{-}(i)}{c_{+}(i)}\\
c_{-}(i) &= \min_{b: b\leq q'} \frac{p_{i,b}}{\pi_b}\\
c_{+}(i) &= \max_{b: b \leq q'} \frac{p_{i,b}}{\pi_b}\,.
\end{align}

\begin{align}
\bbE \eqstartshort \left[\log \bbP[\tilde A=1|\tilde Y ^n = Y^n] \bigg| A=1\right] 
\eqbreakshort&= \log \bbP[N_1 = n \pi_1,\dots, N_{q'} = n \pi_{q'}] \\
&= \log\frac{\alpha^{q'-1}}{n^{(q'-1)/2}  \sqrt{B}} \\
&= \log\left(\frac{\alpha^{q'-1}}{n^{(q'-1)/2} c_{*}^{\frac{q'-1}{2}}} \left( \frac{\pi_{max}}{\prod_{b} \pi_b}\right)^{1/2} \right)\\
&= \log\left(\frac{\alpha^{q'-1}}{ c_{*}^{\frac{q'-1}{2}}} \left( \frac{n\pi_{max}}{\prod_{b = 1}^{q'} n\pi_b}\right)^{1/2} \right)\\
& = \frac{1}{2} \log n \pi_{max} - \sum_{b:\pi_b > 0 }\frac{1}{2} \log n\pi_b + (q'-1)\log \left(\frac{\alpha}{ \sqrt{c_{*}}}\right)\\
& \leq \frac{1}{2} \log n - \sum_{b:\pi_b > 0 }\frac{1}{2} \log n\pi_b + c' \,.\label{eq::prob_final}
\end{align}
where $c'$ is constant that does not depend on $n$. Importantly, the value of $c'$ also does not depend on $\pi_b$ for any $b$. The quantity $c'$ depends on $c_{*}$, which we can compute with:
\begin{align}
\frac{p_{i,b}}{\pi_b} \eqstartshort = \frac{\bbP[X = b| Y = y_i]}{\pi_b} 
\eqbreakshort \eqstartshort = \frac{\bbP[Y = y_i | X = b] \bbP[X = b]}{\pi_b \bbP[Y = y_i]} = \frac{\bbP[Y = y_i | X = b]}{\bbP[Y = y_i]} 
\eqbreakshort \eqstartshort= \frac{P_{Y|X} (y_i|b)}{{\bbP[Y = y_i]} } 
\end{align}
and
\begin{align}
c_{*} &= \min_{i} \frac{\min_{b}\frac{p_{i,b}}{\pi_b}}{\max_{b}\frac{p_{i,b}}{\pi_b}}\\
&= \min_{i} \frac{\min_{b}\frac{P_{Y|X} (y_i|b)}{{\bbP[Y = y_i]} } }{\max_{b}\frac{P_{Y|X} (y_i|b)}{{\bbP[Y = y_i]} } }\\
&= \min_{y} \frac{\min_{b}P_{Y|X} (y|b) }{\max_{b}P_{Y|X} (y|b)}
\end{align}
So $c_{*}$ only depends on $P_{Y|X}$.

Combining these terms with those from \Cref{lem::prob_a} gives that the expression in \eqref{eq::constant_term_to_compute} is a constant when $P_{Y|X}$ is strictly positive. This constant depends on $q$ and $P_{Y|X}$ but not on $n$ or $\pi_b$ for any $b$.

\end{proof}

\begin{proof}[Proof of \Cref{thm::strict_positive}]
Using \Cref{thm::bounds_uniform_type}  with \Cref{prop::covering_argument} completes the proof for strictly positive DMC.
\end{proof}

\section{Conclusion}

In summary, our work determines the capacity of the noisy permutation channel for the case of a strictly positive DMC matrix. Our main method is to use KL divergence covering on the probability simplex. A key ingredient necessary to complete this proof is our theorem which computes the KL divergence between noisy observations of a sequence sampled from a fixed type class versus noisy observations of an iid sequence. 
We expect this key theorem, which is interesting in its own right, to be applicable to other problems as well. We also determine the capacity of the noisy permutation channel for block diagonal DMC matrices with strictly positive blocks, the $q$-ary erasure channel, and the Z-channel. 

Finally, we provide some directions for future research. 
\begin{enumerate}
    \item While we can determine the capacity of the noisy permutation channel for strictly positive DMC matrices and a certain subset of non-strictly positive DMC matrices, we do not know how to compute the capacity for general (non-strictly positive) DMC matrices. We know that the achievability bound \eqref{eq::anuran_ach} will apply in the general case, however strategically placed $0$'s in the DMC matrix could possibly increase the capacity above the rate specified in \eqref{eq::anuran_ach}.
    \item Since our converse bound to the capacity is computed using mutual information, this is only a weak converse bound. This leaves open the question of whether we can find strong converse bounds \cite[Section 22.1]{polyanskiy2014lecture}.  
    %
    \item Capacity gives the (asymptotic) leading coefficient of the $\log n$ term in the expansion of $\log M^*(n,\epsilon)$ as $n\to\infty$. Finding the next-order terms and their dependence on $\epsilon$ would be very interesting.
\end{enumerate}

\appendices

\section{Comparing \Cref{thm::bounds_uniform_type} to Stam}

\label{sec::comparing_stam}

Here we give some details on how our result compares to that of \cite{stam1978}, which we will refer to as Stam's setting or Stam's result. 
Though similar, our setting is not exactly the same as Stam's setting. The differences are:
\begin{enumerate}
    \item Stam's result generalizes to $m$ observations, where $m$ can be less than $n$. Our result \Cref{thm::bounds_uniform_type} only applies to exactly $n$ observations. \label{item::stam_m}
    \item Stam's setting has noiseless observations whereas our setting has noisy observations. \label{item::stam_noiseless}
\end{enumerate}

In order for our result and Stam's result to be comparable, we apply additional theorems to both our result and Stam's result so that we are in a setting where both results are for all $m \leq n$ and for noisy observations. 

Regarding difference \ref{item::stam_m}), we can use a version of Han's inequality for divergence \cite[Proposition 5.5]{duchi2021lecture} (applies when the second probability argument is independent over the entries of the vector $Y^n$) on our result \Cref{thm::bounds_uniform_type}, to get the following corollary:

\begin{fcorollary}\label{cor::m_observations}
Let $Y^n \sim P_{Y|X}^n \circ U$, so that $P_{Y^n}$ is the distribution as in Theorem 2. Then for every $m\le n$ we have:
\begin{align}
D(P_{Y^m} \| Q_{Y}^m) \leq m D(P_{Y} \| Q_{Y}) + \frac{m}{n}c
\end{align}
or when $Q_Y = P_Y$,
\begin{align}\label{eq::m_observations}
D(P_{Y^m} \| P_{Y}^m) \leq \frac{m}{n}c\,.
\end{align}
where $Y^m$ are the first $m$ entries of vector $Y^n$ and $c$ is the same constant as in \Cref{thm::bounds_uniform_type}.
\end{fcorollary}

Regarding difference \ref{item::stam_noiseless}), using data processing inequality, we can use Stam's result as an upper bound for the case with noisy observations. Stam's result with data processing gives
\begin{align}
D \eqstartshort (P_{Y^m} \| P_{Y}^m) \leq D(P_{X^m} \| P_{X}^m) \eqlinebreakshort \leq \frac{(q-1)}{2} \frac{m(m-1)}{(n-1)(n-m+1)}\,.\label{eq::stam_data_process}
\end{align}
The above equation, which is presented as the final result in \cite{stam1978}, is actually not the tightest when $m$ is close to $n$. For instance, when $m = n$, \eqref{eq::stam_data_process} gives $\frac{q-1}{2}n$ which is far from $\frac{q-1}{2}(\log n + c')$, the actual divergence when computed directly. An improvement on Stam's bound when $m$ is close to $n$ is given in \cite{matus2017}. We show an easier improvement, using an intermediate result in the proof of \eqref{eq::stam_data_process}. We can derive for larger $m$ that
\begin{align}\label{eq::stam_data_process_large_m}
D\eqstartshort (P_{Y^m} \| P_{Y}^m) \eqbreakshort&\leq D(P_{X^m} \| P_{X}^m) \\
&\leq \frac{q-1}{n-1} \sum_{t = 1}^{m-1} \frac{t}{n-t} \\
&= \frac{q-1}{n-1} \sum_{j = n - m + 1}^{n-1} \frac{n-j}{j} \\
& = \frac{q-1}{n-1} \left(n\left(\sum_{j = n - m + 1}^{n-1} \frac{1}{j} \right) -(m-1)\right)\\
& = \frac{q-1}{n-1} \left(n\left(\log(n-1) - \log(n-m) + c''\right) -(m-1)\right)\\
& = \frac{q-1}{n-1} \left(n\log\frac{n-1}{n-m} + nc'' -(m-1)\right)\\
& = (q-1) \log \frac{n-1}{n-m} + O(q)\label{eq::stam_data_process_large_m}
\end{align}
for $m < n$ and $c''$ is a constant leftover from approximating the harmonic sum by a logarithm.

We now can compare our result \eqref{eq::m_observations} with Stam's result, either \eqref{eq::stam_data_process} and \eqref{eq::stam_data_process_large_m}, in the setting of $m \leq n$ and noisy observations:
\begin{itemize}
\item When $m << n$, \eqref{eq::stam_data_process} is a better bound than \eqref{eq::m_observations}.
\item When $m$ is very close to $n$, such as when $n-m = o(n)$, \eqref{eq::m_observations} is a tighter bound than both \eqref{eq::stam_data_process} and  \eqref{eq::stam_data_process_large_m}.
\item When $m$ is linear in $n$, then it becomes important to compare the constant factors. Let $\gamma = m/n$. To get an estimate on when our bound is tighter, we first assume $n$ is large and ignore the lower order constants which appear in the bounds. Using \Cref{rem::constant_c}, \eqref{eq::m_observations} is tighter than \eqref{eq::stam_data_process} and  \eqref{eq::stam_data_process_large_m} for large $n$ if %
\begin{align}
\frac{1}{2} \log \frac{2 \pi \alpha^2}{c_*} \leq \min\left\{\frac{\gamma}{1 - \gamma} ,\frac{1}{\gamma}\log \frac{1}{1 - \gamma}\right\}\,.
\end{align}
This can occur for certain values of $\gamma$ depending on the size of $c_*$, which is a function of $P_{Y|X}$.
\end{itemize}
The observations indicate that whether our result is tighter or Stam's result is tighter depends on the value of $m$ and $n$. This also verifies that our \Cref{thm::bounds_uniform_type} result cannot be proven as just a corollary of Stam's result and indeed we are offering something new. Our result can have consequences in Stam's setting for noisy observations when $n$ and $m$ are large.

\section{Covering Results}
\label{sec::covering_results}

In this section we give the covering results necessary for proving \Cref{prop::covering_argument}, which include the proofs of \Cref{thm::covering_non_random} and \Cref{prop::cover_subset}. Again, the bound in \Cref{thm::covering_non_random} is sufficient but not the best possible covering bound. Other bounds are explored in \cite{phdthesis}.

%

\subsection{Divergence Covering Upper Bound (Proof of \Cref{thm::covering_non_random})}

We need some preliminaries before proving \Cref{thm::covering_non_random}.
To define our covering centers for the simplex, we start with a set of scalars. 
Let 
\begin{align}
\Lambda \left(\varepsilon \right) & \eqdef \left\{  {\varepsilon i^2}: \text{ for } i \in \bbZ_{> 0}, {\varepsilon i^2} < \frac{1}{2}  \right\} 
\eqlinebreakshort\cup \left\{ 1 - {\varepsilon i^2}: \text{ for } i \in \bbZ_{> 0}, {\varepsilon i^2} < \frac{1}{2}  \right\}  \cup \left\{ \frac{1}{2}\right\}
\end{align}
Define
\begin{align}
\Lambda_2 \left(\varepsilon \right) = \left\{ (\lambda, 1 - \lambda): \lambda \in \Lambda\left({\varepsilon}\right) \right\}  
\end{align}
For each $k$, let $u_k \in \Delta_{k-1}$ be $u_k = (0, \dots,0, 1)$.
For each $q \in \Delta_{k-2}$, let $\hat q$ be the corresponding $\hat q \in \Delta_{k-1}$ where $\hat q = (q_1,\dots, q_{k-1}, 0) $.

For each $q\in \Delta_{k-2}$, define $q^{(\lambda)}$ such that
\begin{align}
q^{(\lambda)} = \lambda u_k + (1-\lambda) \hat q\,.
\end{align}
For $k>2$, recursively define
\begin{align}
\Lambda_k\left(\varepsilon \right) \eqdef \bigcup_{\lambda \in \Lambda\left(\frac{\varepsilon}{k} \right)} \left\{ q^{(\lambda)}: q \in \Lambda_{k-1}\left(\frac{k-1}{k} \varepsilon\right)  \right\}
\end{align}

\newcommand{\covconst}{\ensuremath{\gamma}}

\begin{flemma}\label{lem::div_array_bound}
For any $p \in \Delta_{k-1}$,
\begin{align}
\min_{q \in \Lambda_k(\varepsilon)}D(p \| q) \leq \covconst \varepsilon
\end{align}
where $\covconst$ is a constant. 
\end{flemma}

\begin{proof}
We show this by using induction. First, for any $p \in \Delta_{1}$, we want to show that
\begin{align}
\min_{q \in \Lambda_2(\varepsilon)}D(p \| q) \leq \covconst \varepsilon\,.
\end{align}
 We use the following fact\footnote{This fact is that KL divergence is upper-bounded by $\chi^2$-divergence.} from \cite{CsiszarSomethingLemma}.
For probabilities $P_1$ and $P_2$ on $\az$ symbols, we have
\begin{align}
D(P_1 || P_2) \leq \sum_{a = 1}^\az \frac{(P_1(a) - P_2(a))^2}{P_2(a)}\,.\label{eq::reversePinsker}
\end{align}

This implies that for any $p, q \in \Delta_1$, where $p=(p_1, 1 -p_1)$ and $q=(q_1, 1 - q_1)$
\begin{align}
D(p || q) \leq \frac{(p_1 - q_1)^2}{q_1} + \frac{(1 -p_1 - 1 + q_1)^2}{1 - q_1} = \frac{(p_1 - q_1)^2}{q_1(1- q_1)}\,.
\end{align}

Suppose that $p \in \Delta_1$ and $p_1 < 1/2$. Then $\varepsilon (i-1)^2 < p_1 \leq \varepsilon i^2$ for some positive integer $i$. Assume for now that $\varepsilon i^2 < 1/2$. Choose $q = (\varepsilon i^2, 1 - \varepsilon i^2) \in \Lambda_2(\varepsilon)$. Note that we must have $1 - \varepsilon i^2 > 1/2$.
\begin{align}
D(p \| q)  &\leq \frac{(p_1 - q_1)^2}{q_1(1-q_1)} \\
&= \frac{(p_1 - \varepsilon i^2)^2}{(\varepsilon i^2)(1- \varepsilon i^2) }\\
&\leq \frac{(\varepsilon(i-1)^2 - \varepsilon i^2)^2}{(\varepsilon i^2)(1- \varepsilon i^2) } \\
& \leq \varepsilon \frac{(-2i + 1)^2}{i^2(1/2)} \\
& \leq \varepsilon \frac{4i^2 - 4i + 1}{i^2/2} \\
& \leq 8 \varepsilon
\end{align}

If $\varepsilon i^2 > 1/2$, we can choose $q = (1/2, 1/2)$. For this case, we can also assume $i > 1$, otherwise one center point, $q = (1/2, 1/2)$ is sufficient for covering the whole simplex. Then

\begin{align}
D(p \| q)  &\leq \frac{(p_1 - 1/2)^2}{(1/2)(1/2)} \\
&\leq \frac{(\varepsilon(i-1)^2 - \varepsilon i^2)^2}{1/4} \\
& \leq 4 \varepsilon^2 {(-2i + 1)^2} \\
& \leq \frac{4}{2} \varepsilon \frac{4i^2 - 4i + 1}{(i-1)^2} \\
& \leq 18 \varepsilon
\end{align}
where we used that $\varepsilon < 1/(2 (i-1)^2)$. This shows that we can set $\covconst = 18$.
By symmetry, $\min_{q \in \Lambda_2(\varepsilon)}D(p\|q) \leq \covconst \varepsilon$ holds for $p_1 > 1/2$ as well.

Suppose in dimension $k-1$, that we have for any $p \in \Delta_{k-2}$, 
\begin{align}
\min_{q \in \Lambda_{k-1}(\varepsilon)}D(p \| q) \leq \covconst\varepsilon
\end{align}
For each $p = (p_1,...,p_k) \in \Delta_{k-1}$, we specify a scalar quantity $\lambda_p \in [0, 1]$. If $p_k <1/2$, like above, we can find a positive integer $i$ where 
\begin{align}
\frac{\varepsilon}{k} (i-1)^2 \leq p_k \leq \frac{\varepsilon}{k} i^2
\end{align}
and set 
\begin{align}
\lambda_p = \min\left\{\frac{\varepsilon}{k} i^2, \frac{1}{2} \right\}\in \Lambda\left(\frac{\varepsilon}{k}\right)\,.
\end{align}

If $p_k  > 1/2$, find $i$ such that 
\begin{align}
 1- \frac{\varepsilon}{k} i^2 < p_k \leq 1- \frac{\varepsilon}{k} (i-1)^2 
\end{align}
 and set 
\begin{align}
 \lambda_p = \max \left\{ 1 -  \frac{\varepsilon}{k} i^2, \frac{1}{2} \right\}\in \Lambda\left(\frac{\varepsilon}{k}\right)\,.
 \end{align}
Define $p_k' = (p_k, 1- p_k)$ and $\lambda_p' = (\lambda_p, 1-\lambda_p)$, then similar to above
\begin{align}
D(p_k' \|\lambda_p') \leq \covconst \frac{\varepsilon}{k}\,.
\end{align}
\begin{align}
\min_{q \in \Lambda_{k}(\varepsilon)} \eqstartshort D(p \| q) \eqbreakshort &\leq \min_{q \in \Lambda_{k}(\varepsilon): q_k = \lambda_p} p_k\log\frac{p_k}{q_k} + \sum_{i = 1}^{k-1} p_i \log \frac{p_i}{q_i} \\
& \leq p_k \log \frac{p_k}{\lambda_p} + \min_{q \in \Lambda_{k}(\varepsilon): q_k = \lambda_p}\sum_{i = 1}^{k-1} p_i \log \frac{p_i}{q_i} \\
& \leq p_k \log \frac{p_k}{\lambda_p} +  \min_{q \in \Lambda_{k}(\varepsilon): q_k = \lambda_p} (1-p_k) \log \frac{1-p_k}{1-\lambda_p} \eqlinebreakshort+ (1-p_k)\sum_{i = 1}^{k-1} \frac{p_i}{1 - p_k} \log \frac{p_i / (1- p_k)}{q_i / (1- \lambda_p)} \\
& \leq D(p_k' \| \lambda_p') +  (1-p_k) \eqlinebreakshort\min_{q' \in \Lambda_{k-1}\left(\frac{k-1}{k}\varepsilon\right)}\sum_{i = 1}^{k-1} \frac{p_i}{1 - p_k} \log \frac{p_i / (1- p_k)}{q'_i} \\
&\leq \covconst\frac{\varepsilon}{k} +  (1-p_k) \covconst \frac{k-1}{k}\varepsilon\\
&\leq \covconst \varepsilon\,.
\end{align}


\end{proof}

\begin{proof}[Proof of \Cref{thm::covering_non_random}]
We use $\cQ_k (\varepsilon)$ to denote the set of centers we need to cover $\Delta_{k-1}$ with radius $\varepsilon$. 

Let 
\begin{align}
\cQ_k(\varepsilon) = \Lambda_k\left(\frac{\varepsilon}{\covconst} \right)
\end{align}
where $\covconst$ is the constant in \Cref{lem::div_array_bound}.
Then using \Cref{lem::div_array_bound}, for $p \in \Delta_{k-1}$, 
\begin{align}
\min_{q \in \cQ_k(\varepsilon)}D(p || q) \leq \varepsilon\,.
\end{align}
Since, $M(k, \varepsilon) \leq |\cQ_k(\varepsilon)|$, it remains to count the size of each $\cQ_k(\varepsilon)$. We show its size by induction.
First, we have that 
\begin{align}
\left|\Lambda\left(\frac{\varepsilon}{\covconst} \right)\right| \leq 2\sqrt{\frac{\covconst}{2\varepsilon}} + 1 = \sqrt{\frac{2\covconst}{\varepsilon}} + 1 \leq \frac{\sqrt{2\gamma} + 1}{\sqrt{\varepsilon}}
\end{align}
where the last inequality holds if $\varepsilon \leq 1$. Therefore we have for some constant $c$ (we can show $c \leq 7$),
\begin{align}
|\cQ_2(\varepsilon) |\leq  c\frac{1}{\sqrt{\varepsilon}}\,.
\end{align}

For the inductive case, given alphabet size $k$ and any $\varepsilon \leq 1$, we have $| \Lambda_k \left(\frac{\varepsilon}{\covconst} \right)|\leq c^{k-1}\left( \frac{k-1}{\varepsilon}\right)^{\frac{k-1}{2}}$.

Now consider the case of alphabet size $k+1$. The set $\Lambda_{k+1}\left(\frac{\varepsilon}{\covconst}\right)$ is defined as a set of points which is a product of sets  $\Lambda \left(\frac{1}{k}\frac{\varepsilon}{\covconst}\right)$ and $\Lambda_{k+1}\left(\frac{k-1}{k}\frac{\varepsilon}{\covconst}\right)$.
This gives
\begin{align}
|\cQ_{k+1}(\varepsilon)| \eqstartshort = \left|\Lambda_{k+1} \left(\frac{\varepsilon}{\covconst}\right) \right| \eqbreakshort
&=  \left|\Lambda \left(\frac{1}{k}\frac{\varepsilon}{\covconst}\right)\right| \left|\Lambda _{k}\left(\frac{k-1}{k}\frac{\varepsilon}{\covconst}\right)\right| \\
&\leq \left(c \frac{1}{\sqrt{\frac{\varepsilon}{k}}} \right) \left(c^{k-1}\left( \frac{k-1}{\varepsilon \frac{k-1}{k}}\right)^{\frac{k-1}{2}}\right)\\
&= c \frac{\sqrt{k}}{\sqrt{\varepsilon} }c^{k-1}\left(\frac{k}{\varepsilon}\right)^{\frac{k-1}{2}}\\
& = c^k\left(\frac{k}{\varepsilon}\right)^{\frac{k}{2}}
\end{align}
as the number of centers. 
\end{proof}



\subsection{Subspace Covering (Proof of \Cref{prop::cover_subset})}

To use our covering result for noisy permutation channels, we actually need to cover a lower dimensional subspace of a $(\az-1)$-dimensional simplex. 

\begin{flemma}\label{lem::map_subspace_covering}
For $\cB \subset \Delta_{\az-1}$, suppose there is a stochastic matrix $F$ which maps $\Delta_{\ell - 1}$ onto $\cB$. Then,
\begin{align}
M(\az, \varepsilon, \cB) \leq M(\ell, \varepsilon)\,.
\end{align}
\end{flemma}

\begin{proof}
Let $\mathcal{N}_c(\ell, \varepsilon)$ be the set of points which are centers for a divergence covering of $\Delta_{\ell - 1}$ with covering radius $\varepsilon$. 
For each $b \in \cB$, there exists a $p \in \Delta_{\ell - 1}$ such that $p F = b$. 
For this $p$, let $r \in \mathcal{N}_c(\ell, \varepsilon)$ be such that $D(p||r) \leq \varepsilon$. Let $b^{*} = rF$. 
By data processing inequality \cite[Theorem 2.2]{polyanskiy2014lecture},
\begin{align}
D(b|| b^*) \leq D(p||r)\leq \varepsilon\,.
\end{align}

Hence the image of the set of centers in $\mathcal{N}_c(\ell, \varepsilon)$ mapped using $F$, becomes the set of centers for a divergence covering on $\cB$ with radius $ \varepsilon$.

\end{proof}

\begin{proof}[Proof of \Cref{prop::cover_subset}]
The key to this proof is to divide the space $\cB$ into simplices of dimension $\ell - 1$.

We can upper bound the number of simplices needed for a partition of $\cB$. The image of $F$ is a convex hull of at most $q$ points (recall $q$ is the size of the input symbols). We call these corner points. Consider all possible choices of $\ell$ of these $q$ corner points. Let this set of all combinations be $S$, where
\begin{align}
|S| = \binom{q}{\ell}\,.
\end{align} 

For each $s \in S$, let $\cB_s$ be the simplex which is the convex hull of the $\ell$ corner points in set $s$. 

For each point $x$ in the image of $F$, since $F$ has rank $\ell$, there exists some linear combination of $\ell$ corner points which results in $x$. If $s$ is this set of $\ell$ points, then $x \in \cB_s$. Thus for all $x \in \cB$, there exists some $s\in S$, so that $x \in \cB_s$.

Label each of these simplices as $\cB_1,...,\cB_{|S|}$. 
There exists a stochastic matrix $F_i$ which maps from space $\Delta_{\ell - 1}$ onto the space $\cB_i$. In particular, we can find this map $F_i$ by mapping each of the $\ell$ corners of $\Delta_{\ell - 1}$ into one of the $\ell$ corner points of $\cB_i$.  This map covers all of $\cB_i$ by linearity. 

Hence using \Cref{lem::map_subspace_covering}, we can find a divergence covering of size $M(\ell, \varepsilon)$ for each $\cB_i$. Combining these covering centers together for all $i$, we get a covering of size 
\begin{align}
\binom{q}{\ell} M(\ell, \varepsilon)\,.
\end{align}

\end{proof}

We are most assuredly over counting the number of simplices $\cB$ has to be divided up into. However, this number does not depend on $\varepsilon$, which is sufficient for our application to noisy permutation channels.

\section{Block Diagonal Case}

\label{sec::block_diagonal}

With a small modification to the proof of \Cref{thm::strict_positive}, we can show a converse bound for block diagonal matrices where each block is strictly positive. The key idea is that since each block is independent from all the other blocks, so we can apply the bound for strictly positive matrices separately to each block. We need to show a separate achievability result to match this converse bound.

As a sanity check, the block diagonal case captures the situation where $P_{Z|X}$ is the identity matrix. In which case, it is possible to use all possible permutations of symbols as messages. No errors are allowed so decoding is straight-forward. Using an identity matrix of size $q\times q$ for the DMC, for each $n$,

\begin{align}
R \eqstartshort = \frac{\log M}{\log n} \eqbreakshort \eqstartshort = \frac{\log \binom{n}{q-1}}{\log n}  \eqbreakshort \eqstartshort \approx \frac{\log (c^{q-1}n^{q-1} / (q-1)^{q-1})}{\log n}  \eqbreakshort \eqstartshort = q-1 + \frac{\log (c^{q-1} / (q-1)^{q-1})}{\log n}
\end{align}
which goes to $q-1$ asymptotically as $n$ increases. This matches our block diagonal converse result.

\subsection{Converse}

\label{sec::block_diagonal_converse}

\begin{fproposition}[Block Diagonal Converse]\label{prop::block_diagonal_converse}
Suppose $P_{Z|X}$ can be written as a block diagonal matrix with $\beta$ blocks, so that each block is strictly positive. 
Then,
\begin{align}
\permcap(P_{Z|X}) \leq \frac{\rank(P_{Z|X}) + \beta - 2}{2}\,.
\end{align}
\end{fproposition}

\begin{proof}

We want to use \Cref{prop::covering_argument} but we need to show a version of the upper bound in \Cref{thm::bounds_uniform_type} which applies to block diagonal matrices instead of strictly positive matrices. 

Fix $\pi \in \cP_n$. Arrange the matrix $P_{Z|X}$ in block diagonal form and let $\cX_b$ be the set of symbols in $\cX$ which are in the $b$th block. Let $(X, Y)^n$ be generated iid from $(\pi \times P_{Y|X})$. Let $W_i$ be the number of $X$ which equals $i$, i.e. 
\begin{align}
W_i = |\{t: X_t = i\}|\,.
\end{align}
Define 
\begin{align}
A_b = \left\{ \bigcap_{i \in  \cX_b }W_i = \pi_i n \right\}\,.
\end{align}
This is the event that all symbols $i$ associated with block $b$ occur with the count $\pi_i n$. Each block has its own separate set of output symbols in $\cY$. The probability of $W_i$ is independent of what happens in other blocks. Let $Y^n(b)$ (and $y^n(b)$) be notation for the symbol counts restricted to just the output symbols associated with the $b$th block. 

Using the definition in \Cref{prop::compute_divergence}, notice that $\bbI[A = 1] = \bbI\left[\bigcap_{b = 1}^\beta A_b\right]$. (Recall the notation $\tilde A = \{\tilde X^n \in T(P) \}$ where $(\tilde X, \tilde Y)^n$ is independent copy under the same distribution $(P \times P_{Y|X})$ as $(X, Y)^n$). Then using \eqref{eq::divergence_diff_bound} with \Cref{lem::prob_a}, 
\begin{align}
D \eqstartshort ( P_{Y|X}^n \circ U \| Q_Y^n) \eqbreakshort
&= nD(P_{Y} \| Q_Y) + \bbE\left[\log \frac{\bbP[\tilde A = 1|\tilde Y^n = Y^n]}{\bbP[\tilde A=1]} \bigg| A=1\right] \\
& \leq nD(P_{Y} \| Q_Y) - \frac{1}{2} \log n + \sum_{i: \pi_i > 0 } \frac{1}{2} \log \pi_i n \eqlinebreakshort + c + \bbE\left[\log \bbP[\tilde A = 1|\tilde Y^n = Y^n] \bigg| A=1\right] \,.
\end{align}

For any $Y^n$,
\begin{align}
\bbP [A = 1 | Y^n ] = \prod_{b = 1}^\beta \bbP [A_b = 1 | Y^n(b) =  y^n(b)]\,.
\end{align}
Each block is a strictly positive matrix. From \Cref{lem::balls_in_bins} and following the same calculations that results in \eqref{eq::prob_final}, we know that 
\begin{align}
 \log \eqstartshort \bbP [A_b = 1 | Y^n(b) =  y^n(b)] \eqbreakshort & \leq \frac{1}{2} \log n - \sum_{i \in \cX_b:\pi_i > 0 }\frac{1}{2} \log n\pi_i + c'
\end{align}
and thus
\begin{align}
\log \eqstartshort\bbP [A = 1 | Y^n ] \eqbreakshort
&\leq \sum_{b = 1}^{\beta} \left(\frac{1}{2} \log n - \sum_{i \in \cX_b:\pi_i > 0 }\frac{1}{2} \log n\pi_i + c' \right)\\
& = \frac{\beta}{2} \log n - \sum_{i :\pi_i > 0 }\frac{1}{2} \log n\pi_i + \beta c'\,.
\end{align}
This holds for all $Y^n$ so it automatically gives the expected value. 
Putting all these terms together, for any $\pi$, we get
\begin{align}
D( P_{Y|X}^n \circ U \| Q_Y^n) & = n D(P_Y|| Q_Y) + \frac{\beta - 1}{2} \log n + c'' 
\end{align}
where $c''$ combines all the constants. 
Using \Cref{prop::covering_argument}, gives
\begin{align}
\permcap(P_{Z|X}) &\leq \frac{\rank(P_{Z|X}) - 1}{2} + \lim_{n \to \infty}\frac{\frac{\beta - 1}{2} \log n + c''}{\log n}\\
& = \frac{\rank(P_{Z|X}) - 1}{2} + \frac{\beta - 1}{2} \\
& = \frac{\rank(P_{Z|X}) + \beta - 2}{2}
\,.
\end{align}

\end{proof}

\subsection{Achievability}

\label{sec::block_diagonal_achievability}

\begin{fproposition}[Block Diagonal Achievability]\label{prop::block_diagonal_achievability}

Suppose $P_{Z|X}$ can be written as a block diagonal matrix with $\beta$ blocks, so that each block is strictly positive. 
Then, 
\begin{align}
\permcap(P_{Z|X}) \geq \frac{\rank(P_{Z|X}) + \beta - 2}{2}\,.
\end{align}
\end{fproposition}

\begin{proof}

The achievability proof encodes using two steps. The first step is a zero-error code based on which block in the block diagonal matrix the symbols are associated with. Let $M_1$ denote the total possible messages (or rather message stems) for the first step. The second step operates only on each block independently, and uses the achievability given by \eqref{eq::anuran_ach}. Let $M_2$ denote the total messages (or message tails) possible here.

Label the $\beta$ blocks in $P_{Z|X}$ as $B_1,...,B_{\beta}$. Define the sets of input symbols $\cX_1,..,\cX_{\beta}$ and output symbols $\cY_1,...,\cY_{\beta}$, so that $\cX_b$ and $\cY_b$ are the input and output symbols respectively associated with block $B_b$. (In other words, if $p_{ij} > 0$ and $p_{ij}$ falls into block $B_b$, then $i \in \cX_b$ and $j \in \cY_b$.) These sets $\cX_1,..,\cX_{\beta}$ and $\cY_1,...,\cY_{\beta}$ are both disjoint. 

Let $L = \rank(P_{Z|X})$ and let $L_b = \rank(B_b)$. Because of the block diagonal structure, $L = \sum_{b = 1}^\beta L_b$.

For fixed $n$, set aside the first $n/2$ input symbol positions so that exactly $n/(2\beta)$ are from set $\cX_b$ for each $b$. These are not used for the first step of the two-step code and are used to make the analysis of the second step easier. 
The remaining $n/2$ positions can be encoded using symbols from any set and this is used to make the first step of the code. There are 
\begin{align} \label{eq::messages_block}
\binom{n/2}{\beta - 1} \geq \left(\frac{n/2}{\beta - 1} \right)^{\beta - 1}
\end{align}
possible combinations of symbols chosen from $\beta$ blocks, disregarding order. The DMC maps the symbols in set $\cX_b$ to symbols in set $\cY_b$ without any error. Hence, \eqref{eq::messages_block} is the number of messages $M_1$ the first step can encode without any error.   

Once it is determined how many symbols of each set will be used, we can determine which symbol in the set will be used for the second step. Suppose there are $n_b$ positions which are designated for symbols in set $\cX_b$. This includes the $n/(2\beta)$ we set aside in the beginning and how ever many were chosen to make the first step of the code. Using \eqref{eq::anuran_ach}, we know there exists a encoder-decoder pair $(f_{n_b}, g_{n_b})$ so that the decoding error is vanishingly small as $n_b \to \infty$. Just by choosing which symbol in $\cX_b$ to send, for some $\varepsilon_{n_b} > 0$ where $\varepsilon_{n_b} \to 0$, we can encode a set of messages with size $M_b$ satisfying 
\begin{align}
\log M_b \geq {\left(\frac{L_b - 1}{2} - \varepsilon_{n_b}\right) \log n_b}\,.
\end{align}

The set of messages possible for all the $\beta$ different sets is
\begin{align}
\log M_2 &= \log \prod_{b = 1}^{\beta} M_b \\
&\geq \sum_{b = 1}^{\beta}{\left(\frac{L_b - 1}{2} - \varepsilon_{n_b}\right) \log n_b} \\
&\geq \sum_{b = 1}^{\beta}{\left(\frac{L_b - 1}{2} - \varepsilon_{n_b}\right) \log \frac{n}{2\beta}} \\
&={\left(\frac{L - \beta}{2} - \sum_{b = 1}^\beta \varepsilon_{n_b}\right) \log \frac{n}{2\beta}}\,.  
\end{align}

The total number of messages is the product of those available at the first and second steps. 
\begin{align}
\log M &= \log M_1 + \log M_2 \\
& \geq \log \left(\frac{n/2}{\beta - 1} \right)^{\beta - 1}+ {\left(\frac{L - \beta}{2} - \sum_{b = 1}^\beta \varepsilon_{n_b}\right) \log \frac{n}{2\beta}}  \\
& = (\beta - 1) \log \frac{n}{2\beta - 2} + \left(\frac{L - \beta}{2} - \sum_{b = 1}^\beta \varepsilon_{n_b}\right)\log  \frac{n}{2\beta}\\
& \geq  \left(\frac{2\beta - 2}{2} + \frac{L - \beta}{2} - \sum_{b = 1}^\beta \varepsilon_{n_b}\right)\log  \frac{n}{2\beta}\\
&= \left(\frac{L + \beta - 2}{2} - \sum_{b = 1}^\beta \varepsilon_{n_b}\right)\log  \frac{n}{2\beta}\,.
\end{align}

Since each $n_b \geq \frac{n}{2\beta} \to \infty$ as $n\to \infty$, asymptotically the term $\sum_{b = 1}^\beta \varepsilon_{n_b}$ disappears. 

The achievable rate is given by 
\begin{align}
R &= \frac{\log M}{\log n} \eqbreakshort \eqstartshort\geq \left(\frac{L - \beta - 2}{2} + \sum_{b = 1}^\beta \varepsilon_{n_b}\right)\frac{\log n - \log 2\beta}{\log n} \to \frac{L - \beta - 2}{2}\,.
\end{align}
\end{proof}

Combining \Cref{prop::block_diagonal_converse} and \Cref{prop::block_diagonal_achievability} gives the first result in \Cref{thm::other_channels}.

\section{Erasure and Z-Channels}\label{sec::other_channels}

\subsection{Concentration Lemma}

The following lemma is useful for computing the probability of $A = 1$ (see \Cref{prop::compute_divergence}) when the DMC matrix is not strictly positive. It is a straight-forward concentration bound which is a direct application of Bernstein's inequality. We choose to write the proof anyways for completeness.

\begin{flemma}\label{lem::concentration_sum}
Suppose that $Z$ is a sum of $n$ independent Bernoulli random variables. Let $\bbE [Z]$ be the expected value of $Z$.

Fix constant $\gamma$. If $\bbE [Z]> 2 \gamma\log n$, then with probability at least $1 - 2/n^{\gamma/4}$, we have that
\begin{align}
\bbE[Z] > \bbE [Z] - \sqrt{\bbE[Z] \gamma\log n} \geq \frac{1}{5} \bbE [Z]\,.
\end{align} 
\end{flemma}

\begin{proof}

Let $Z = \sum_{i = 1}^n W_i$ where $W_i$ is the $i$th Bernoulli random variable. Let $0<p_i<1$ be the probability of $W_i = 1$.
\begin{align}
\sum_{i = 1}^n\bbE[(W_i - \bbE[W_i])^2] = \sum_{i = 1}^n p_i (1 - p_i) \leq \sum_{i = 1}^n p_i = \bbE [Z] \,.
\end{align}
Next, we use Bernstein's inequality for bounded variables \cite[Theorem 2.8.4]{vershynin2018}.

\begin{align}
\bbP \eqstartshort \left[Z - \bbE[Z] \leq -\sqrt{\bbE [Z] \gamma\log n}\right] \eqbreakshort
&\leq 2\exp \left( \frac{-\frac{1}{2} \bbE [Z] \gamma \log n }{\sum_{i = 1}^n\bbE[(W_i - \bbE[W_i])^2]  + \frac{1}{3} 1 \sqrt{\bbE [Z] \gamma\log n}}  \right)\\
& \leq 2\exp \left( \frac{-\frac{1}{2} \bbE[Z] \gamma\log n}{\bbE [Z] + \frac{1}{3} \sqrt{\bbE [Z] \gamma \log n}}  \right)\\
& \leq 2\exp \left( \frac{-\frac{1}{2} \gamma \log n}{1 + \frac{1}{3} \frac{\sqrt{\gamma \log n}} {\sqrt{\bbE Z}}} \right)\,.
\end{align}

Using that $\bbE [Z]> 2 \gamma\log n$, we have $1 + \frac{1}{3} \frac{\sqrt{\gamma \log n}} {\sqrt{\bbE [Z]}} \leq 1 + \frac{1}{3} \frac{\sqrt{ \gamma\log n}} {\sqrt{2\gamma\log n}} \leq 2$. 

\begin{align}
\bbP \eqstartshort \left[Z - \bbE[Z] \leq -\sqrt{\bbE [Z] \gamma \log n}\right] 
\eqbreakshort \eqstartshort \leq 2\exp\left( \frac{-1}{4} \gamma\log n \right) 
\eqbreakshort \eqstartshort \leq \frac{2}{n ^{\gamma/4}}\,.
\end{align}
Hence, with probability $1 - {2}/{n ^{\gamma/4}}$, 
\begin{align}
\bbE[Z] &\geq \bbE [Z] - \sqrt{\bbE Z\gamma\log n}  
\eqbreakshort \eqstartshort\geq \bbE [Z] - \sqrt{\bbE [Z] \frac{1}{2} \bbE [Z]} 
\eqbreakshort \eqstartshort\geq \left(1 - \frac{1}{\sqrt{2}} \right) \bbE [Z] 
\eqbreakshort \eqstartshort\geq \frac{1}{5} \bbE [Z]\,.
\end{align}

\end{proof}

\subsection{The $q$-ary Erasure Channel}
We now can prove the converse bound for $q$-ary erasure channels, where $q$ is the number of input symbols.  Let $\az = q+1$ represent the erased symbol. 

The matrix $P_{Z|X}$ for a $q$-ary erasure channel has the following structure:
\begin{align}
P_{Z|X} = 
\begin{bmatrix}
p_{11} & 0  & \cdots & 0 & p_{1\az} \\
0 & p_{22}  & \cdots & 0 & p_{2\az} \\
\vdots & \vdots  & \ddots & \vdots & \vdots \\
0 & 0  & \cdots & p_{qq} & p_{q\az} \\
\end{bmatrix}\,.
\end{align}
We assume that $p_{i\az} > 0$ for each $i$. 

\begin{proof}[Proof of \Cref{cor::q_erasure} of \Cref{thm::other_channels}]

Fix $\pi = (\pi_1,...,\pi_q)$ where $\pi \in \cP_n$. (We assume each $\pi_i >0$, otherwise we can remove it.) Reorder the symbols in $\{1,...,q\}$ so that $\pi_1 \leq \pi_2 \leq ... \leq \pi_q$. 
(Note that $P_{Y|X} = P_{Z|X}$.) 

Following \Cref{prop::compute_divergence}, let $(X, Y)^n$ be generated iid according to $(\pi \times P_{Y|X})$. To use \Cref{prop::compute_divergence} we need to determine $\bbP[\log \bbP[\tilde A = 1|\tilde Y^n = Y^n] | A = 1]$.

Unlike the case of strictly positive $P_{Y|X}$, the value of $\bbP[A = 1 | Y^n]$ depends on $Y^n$. For instance, it is easy to see that when the erasure symbol $\az$ does not appear, then $\bbP[A = 1 | Y^n] = 1$. While $Y^n$ like this can occur under the event $A = 1$, we want to show that these events are rare, this way the expected value of $\bbP[A = 1 | Y^n]$ given that $A = 1$ is much smaller than $1$ and close to the value which will give our result. We first show a concentration result on $Y^n$ given that $A = 1$. 

Let $U_{b}$ be the random variable which gives the count of the number of times the symbol $b$ is erased, i.e
\begin{align}
U_b = \sum_{i = 1}^n \bbI\{(X_i, Y_i) = (b, \az) \}\,.
\end{align}

Let $v_b(y^n) = \{ U_{b}| A = 1, Y^n = y^n \}$. Note that $v_b(y^n)$ is deterministic. If $A = 1$ and $Y^n$ is known, we can determine exactly what $U_b$ is.  

Define $S_b = \sum_{a \geq b} U_a$. Given $Y^n$, $S_1$ is deterministic. Given $Y^n$ and $U_1,..,U_{b-1}$, $S_b$ is deterministic. 

 Using \Cref{lem::concentration_sum}, since $\bbE[S_b] = n \sum_{a \geq b} \pi_a p_{a \az} \geq n \pi_q p_{q \az} > 2 \gamma\log n$ for some $\gamma$ (chosen later) and all $b$ for large enough $n$, we have
\begin{align}
\bbP\left[S_b > \frac{1}{5} n \sum_{a \geq b} \pi_a p_{a\az}\right] \geq 1 - 2/n^{\gamma/4}\,.
\end{align} 
Using the union bound, 
\begin{align}\label{eq::concentration_Si}
 \bbP\left[ \bigcap_{b = 1}^q \left\{S_b > \frac{1}{5} n \sum_{a \geq b} \pi_a p_{a\az} \right\}\right] \geq 1 - 2q/n^{\gamma/4}\,.
\end{align} 
Next, for any $y^n$ which has positive probability given $A = 1$, 
\begin{align}
\bbP\eqstartshort[A = 1 | Y^n = y^n] \eqbreakshort
 &= \bbP\left[\bigcap_{b = 1}^q U_b = v_b(y^n) \bigg| Y^n = y^n\right]\\
&= \prod_{b = 1}^{q-1}\bbP\left[ U_b = v_b(y^n) \bigg| \bigcap_{a = 1}^{b-1}  U_a = v_a(y^n) , Y^n = y^n\right]\,.
\end{align}

We compute the following which is like the proof \Cref{lem::balls_in_bins} of but with appropriate adjustments. Using \eqref{eq::variance_proxy}, 
\begin{align}
\bbP \eqstartshort\left[ U_b = v_b(y^n) \bigg| \bigcap_{a = 1}^{b-1}  U_a = v_a(y^n) , Y^n = y^n\right] 
\eqlinebreakshort\leq \frac{\alpha}{\sqrt{ \sum_{i = 1} ^{S_b} \min\left\{\frac{\pi_b p_{b\az}}{\sum_{a \geq b} \pi_a p_{a\az} }, \frac{\sum_{a > b} \pi_a p_{a\az}}{\sum_{a \geq b} \pi_a p_{a\az}} \right\}}}\,.
\end{align}
Like in \Cref{lem::balls_in_bins}, define $c_{-} = \min_{i} p_{i\az}$.
\begin{align}
\min \eqstartshort\left\{\frac{\pi_b p_{b\az}}{\sum_{a \geq b} \pi_a p_{a\az} }, \frac{\sum_{a > b} \pi_a p_{a\az}}{\sum_{a \geq b} \pi_a p_{a\az}} \right\}  
\eqbreakshort & = (\min_{i} p_{i\az}) \min\left\{\frac{\pi_b}{\sum_{a \geq b} \pi_a p_{a\az}  }, \frac{\sum_{a > b} \pi_a }{\sum_{a \geq b} \pi_a p_{a\az} } \right\}\\
& =  \frac{c_{-} \pi_b}{\sum_{a \geq b} \pi_a p_{a\az}}\,.
\end{align}
We get the last equality since $\pi_i$ is in increasing order. 
Hence
\begin{align}
\bbP\eqstartshort \left[ U_b = v_b(y^n) \bigg| \bigcap_{a = 1}^{b-1}  U_a = v_a(y^n) , Y^n = y^n\right]
\eqlinebreakshort \leq \frac{\alpha}{\sqrt{S_b \frac{c_{-} \pi_b}{\sum_{a \geq b} \pi_a p_{a\az}} }}\,.
\end{align}

We can now compute
\begin{align}
\bbE \eqstartshort[\log \bbP[A = 1|Y^n] | A = 1] 
\eqbreakshort &= \sum_{y^n} \bbP[Y^n = y^n | A = 1] \log \bbP[A = 1 | Y^n = y^n]\\
&\leq \log \sum_{y^n} \bbP[Y^n = y^n | A = 1] \bbP[A = 1 | Y^n = y^n]\\
&\leq \log \sum_{y^n} \bbP[Y^n = y^n | A = 1] \prod_{b = 1}^{q-1} \frac{\alpha}{\sqrt{S_b \frac{c_{-} \pi_b}{\sum_{a \geq b} \pi_a p_{a\az}} }}\\
&\leq \log \Bigg( (2q/n^{\gamma/4}) + ( 1 - 2q/n^{\gamma/4})  \eqlinebreakshort\prod_{b = 1}^{q-1} \frac{\alpha}{\sqrt{\left(\frac{1}{5} n \sum_{a \geq b} \pi_a p_{a\az}\right) \frac{c_{-} \pi_b}{\sum_{a \geq b} \pi_a p_{a\az}} }} \Bigg) \label{eq::sub_in_concentration}\\
&\leq \log \left( (2q/n^{\gamma/4}) +  \frac{\alpha^q}{\left(\frac{c_{-}}{5} \right)^{\frac{q - 1}{2}} n^{\frac{q - 1}{2}} \sqrt{\frac{\prod_{b = 1}^{q} \pi_b}{\pi_{max}} } } \right)
\end{align}
where in \eqref{eq::sub_in_concentration} we used \eqref{eq::concentration_Si}. 
We can pick $\gamma$ large enough\footnote{For instance, we can pick $\gamma = 40q$. For large enough $n$, we still get that $\bbE[S_b] = n \sum_{a \geq b} \pi_a p_{a\az} > 2 \gamma\log n$ is true for all $b$.} so that the first term in the logarithm is negligible compared to the second term for large $n$.

This gives
\begin{align}
\bbP\eqstartshort[\log \bbP[\tilde A = 1|\tilde Y^n = Y^n] | A = 1] \eqbreakshort
&\leq \log  \left(  \frac{2\alpha^q}{\left(\frac{c_{-}}{5} \right)^{\frac{q - 1}{2}} n^{\frac{q - 1}{2}} \sqrt{\frac{\prod_{b = 1}^{q} \pi_b}{\pi_{max}} } } \right)\\
& \leq \frac{1}{2} \log n - \sum_{b = 1}^q \frac{1}{2} \log \pi_b n + c'\,.
\end{align}
The value $c'$ collects all the constants. Combining with \Cref{lem::prob_a}, we get that for the $q$-ary erasure channel and sufficiently large $n$ that
\begin{align}
D(P_{Y|X}\circ U \| Q_Y^n ) \leq n D(P_Y \| Q_Y) + c
\end{align}
where $c$ does not depend on $n$ or $\pi$. Using \Cref{prop::covering_argument} completes the converse bound for the proof.

The matching achievability bound needed to get the final capacity result is given in \cite{makur_2020}.

\end{proof}



\subsection{Z-Channel}

The matrix for the Z-channel \cite[p 225]{coverthomas} is 

\begin{align}
\begin{bmatrix}
1 & 0\\
p_{21} & p_{22} 
\end{bmatrix}
\end{align}
where we require that $p_{ij} > 0$. (Typically, $p_{21} = p_{22} = 1/2$, but we consider a more general case here.)

We can actually get the capacity of the noisy permutation channel with the Z-channel without altering the proof for the $q$-ary erasure channels. The transition matrix for the Z-channel can be written as 
\begin{align}
\begin{bmatrix}
p_{11} & 0 & p_{13} \\
p_{21} & p_{22} & 0 
\end{bmatrix}
\end{align}
setting $p_{13} = 0$. This does not change the rank of the matrix or the analysis in the proof. 

\begin{fcorollary}\label{prop::z-channel}
Let $P_{Z|X}$ be a stochastic matrix for the Z-channel, then
\begin{align}
C_{\text{perm}}(P_{Z|X}) = \frac{1}{2}\,.
\end{align}
\end{fcorollary}

This is \Cref{cor::z_cap} of \Cref{thm::other_channels}.

\subsection{``Zigzag'' Channel}

In this section, we explore the limits of our approach. We have a particular DMC matrix which is similar to the $q$-ary erasure channel, but our method is not known to give a tight converse. We use a matrix which could be considered a $q$-ary Z-channel and call it a ``zigzag'' channel since its edges in a transition diagram form a zigzag.

The matrix has the form:

\begin{align}
\begin{bmatrix}
p_{11} & p_{12} & 0  & \cdots & 0 & 0 \\
0 & p_{22} & p_{23} & \cdots & 0 & 0\\
0 & 0 & p_{33} & \cdots & 0 & 0\\
\vdots & \vdots & \vdots & \ddots & \vdots & \vdots \\
0 & 0 & 0 & \cdots & p_{q - 1, q - 1} & p_{q-1, q}\\
0 & 0 & 0 & \cdots & 0 & p_{q q}\\
\end{bmatrix}
\end{align}
where each $p_{ij} > 0$. 
This matrix has rank $q$.

Suppose that $q$ is odd and that $\pi$ is such that $\pi_i$ is $0$ for all even values of $i$. Following the notation and method in \Cref{prop::compute_divergence}, $\bbP[A = 1| Y^n] = 1$, since any output symbol can be decoded to exactly one input symbol. 
For any $\pi$ of this choice, 
\begin{align}
D\eqstartshort(P_{Y|X} || Q^n_{Y}) 
\eqbreakshort &=  - nD(P_Y || Q_Y) - \frac{1}{2} \log n + \sum_{i: \pi_i > 0}\frac{1}{2} \log \pi_i n \eqlinebreakshort + c  + \bbE[\log\bbP[A = 1| Y^n]| A = 1]\\
&= - nD(P_Y || Q_Y)  - \frac{1}{2} \log n + \sum_{i: \pi_i > 0}\frac{1}{2} \log \pi_i n + c\\
&\leq - nD(P_Y || Q_Y) - \frac{1}{2} \log n + \frac{q+1}{2}\frac{1}{2} \log n + c\\
&\leq - nD(P_Y || Q_Y) + \frac{q-1}{4} \log n + c\,.
\end{align}

If $\pi$ of this form is the worst case $\pi$ to use, meaning it gives the largest possible value of $D(P_{Y|X} \circ U \| Q_{Y^n})$ for any $Q_Y$, then we get that
\begin{align}
\permcap(P_{Z|X}) \leq \frac{q - 1}{2} + \frac{q-1}{4} = \frac{3(q-1)}{4} \,.\label{eq::zigzag_bound}
\end{align}
If there is another $\pi$ which is the worst, then our upper bound on the capacity is larger than the value on the right-hand side of \eqref{eq::zigzag_bound}. In either case, there is a gap between our upper bound for the capacity and the lower bound of $(q-1)/2$ given by \eqref{eq::anuran_ach}. Exploring this gap is an opportunity for future work.

\section*{Acknowledgement}
\noindent We would like to thank A. Makur for inspiring this project and for his discussions with us.

\bibliographystyle{IEEEbib}
\bibliography{ref_perm_channel}

\begin{IEEEbiographynophoto}{Jennifer Tang}
Jennifer Tang is a Postdoctoral Associate in IDSS at MIT. Jennifer received her Ph.D. in the Department of Electrical Engineering and Computer Science at MIT while working in LIDS. Previous to that, she received her B.S.E. in Electrical Engineering at Princeton University. Her research interests include information theory, prediction and learning theory, quantization and data compression, high-dimensional statistics, data analytics, defect tolerance, and models for social dynamics and inference. 
\end{IEEEbiographynophoto}

\begin{IEEEbiographynophoto}{Yury Polyanskiy}
Yury Polyanskiy is a Professor of Electrical Engineering and Computer Science and a member of LIDS, IDSS and Center of Statistics. Yury received his M.S. degree in applied mathematics and physics from the Moscow Institute of Physics and Technology, Moscow, Russia in 2005 and his Ph.D. degree in electrical engineering from Princeton University, Princeton, NJ in 2010. His research interests span information theory, statistical machine learning, error-correcting codes, wireless communication and fault tolerance. Dr. Polyanskiy won the 2020 IEEE Information Theory Society James Massey Award, 2013 NSF CAREER award and 2011 IEEE Information Theory Society Paper Award.
\end{IEEEbiographynophoto}

\end{document}